\documentclass[ALICE,manyauthors]{cernphprep}
\usepackage{float}
\usepackage[comma,square,numbers,sort&compress]{natbib}
\usepackage{hyperref}
\usepackage{lineno}
\usepackage{xspace}
\usepackage[dvipsnames]{xcolor}
\usepackage{amsmath}
\usepackage{multirow}
\usepackage[T1]{fontenc}
\usepackage{orcidlink}

\begin{document}
%

\newcommand{\pp}           {pp\xspace}
\newcommand{\ppbar}        {\mbox{$\mathrm {p\overline{p}}$}\xspace}
\newcommand{\XeXe}         {\mbox{Xe--Xe}\xspace}
\newcommand{\PbPb}         {\mbox{Pb--Pb}\xspace}
\newcommand{\pA}           {\mbox{pA}\xspace}
\newcommand{\pPb}          {\mbox{p--Pb}\xspace}
\newcommand{\AuAu}         {\mbox{Au--Au}\xspace}
\newcommand{\dAu}          {\mbox{d--Au}\xspace}

\newcommand{\s}            {\ensuremath{\sqrt{s}}\xspace}
\newcommand{\snn}          {\ensuremath{\sqrt{s_{\mathrm{NN}}}}\xspace}
\newcommand{\pt}           {\ensuremath{p_{\rm T}}\xspace}
\newcommand{\meanpt}       {$\langle p_{\mathrm{T}}\rangle$\xspace}
\newcommand{\ycms}         {\ensuremath{y_{\rm CMS}}\xspace}
\newcommand{\ylab}         {\ensuremath{y_{\rm lab}}\xspace}
\newcommand{\etarange}[1]  {\mbox{$\left | \eta \right |<#1$}}
\newcommand{\yrange}[1]    {\mbox{$\left | y \right |<#1$}}
\newcommand{\dndy}         {\ensuremath{\mathrm{d}N/\mathrm{d}y}\xspace}
\newcommand{\dndeta}       {\ensuremath{\mathrm{d}N_\mathrm{ch}/\mathrm{d}\eta}\xspace}
\newcommand{\avdndeta}     {\ensuremath{\langle\dndeta\rangle}\xspace}
\newcommand{\dNdy}         {\ensuremath{\mathrm{d}N_\mathrm{ch}/\mathrm{d}y}\xspace}
\newcommand{\Npart}        {\ensuremath{N_\mathrm{part}}\xspace}
\newcommand{\Ncoll}        {\ensuremath{N_\mathrm{coll}}\xspace}
\newcommand{\dEdx}         {\ensuremath{\textrm{d}E/\textrm{d}x}\xspace}
\newcommand{\RpPb}         {\ensuremath{R_{\rm pPb}}\xspace}
\newcommand{\pns}          {\ensuremath{{\rm P}(n_{\rm S})}\xspace}
\newcommand{\inelgz}       {\ensuremath{{\rm INEL>0}}\xspace}
\newcommand{\multdensmid}  {\ensuremath{\langle\dndeta\rangle_{|\eta|~<~0.5}}\xspace}
\newcommand{\dndetamid}    {\ensuremath{\dndeta|_{|\eta|~<~0.5}}\xspace}
\newcommand{\acceff}       {\ensuremath{A\kern-.15em\times\kern-.15em\varepsilon}\xspace}

\newcommand{\nineH}        {$\sqrt{s}~=~0.9$~Te\kern-.1emV\xspace}
\newcommand{\seven}        {$\sqrt{s}~=~7$~Te\kern-.1emV\xspace}
\newcommand{\twoH}         {$\sqrt{s}~=~0.2$~Te\kern-.1emV\xspace}
\newcommand{\twosevensix}  {$\sqrt{s}~=~2.76$~Te\kern-.1emV\xspace}
\newcommand{\five}         {$\sqrt{s}~=~5.02$~Te\kern-.1emV\xspace}
\newcommand{\twosevensixnn}{$\sqrt{s_{\mathrm{NN}}}~=~2.76$~Te\kern-.1emV\xspace}
\newcommand{\fivenn}       {$\sqrt{s_{\mathrm{NN}}}~=~5.02$~Te\kern-.1emV\xspace}
\newcommand{\LT}           {L{\'e}vy-Tsallis\xspace}
\newcommand{\GeVc}         {Ge\kern-.1emV/$c$\xspace}
\newcommand{\MeVc}         {Me\kern-.1emV/$c$\xspace}
\newcommand{\TeV}          {Te\kern-.1emV\xspace}
\newcommand{\GeV}          {Ge\kern-.1emV\xspace}
\newcommand{\MeV}          {Me\kern-.1emV\xspace}
\newcommand{\GeVmass}      {Ge\kern-.2emV/$c^2$\xspace}
\newcommand{\MeVmass}      {Me\kern-.2emV/$c^2$\xspace}
\newcommand{\lumi}         {\ensuremath{\mathcal{L}}\xspace}

\newcommand{\ITS}          {\rm{ITS}\xspace}
\newcommand{\TOF}          {\rm{TOF}\xspace}
\newcommand{\ZDC}          {\rm{ZDC}\xspace}
\newcommand{\ZDCs}         {\rm{ZDCs}\xspace}
\newcommand{\ZNA}          {\rm{ZNA}\xspace}
\newcommand{\ZNC}          {\rm{ZNC}\xspace}
\newcommand{\SPD}          {\rm{SPD}\xspace}
\newcommand{\SDD}          {\rm{SDD}\xspace}
\newcommand{\SSD}          {\rm{SSD}\xspace}
\newcommand{\TPC}          {\rm{TPC}\xspace}
\newcommand{\TRD}          {\rm{TRD}\xspace}
\newcommand{\VZERO}        {\rm{V0}\xspace}
\newcommand{\VZEROA}       {\rm{V0A}\xspace}
\newcommand{\VZEROC}       {\rm{V0C}\xspace}
\newcommand{\Vdecay} 	   {\ensuremath{V^{0}}\xspace}

\newcommand{\ee}           {\ensuremath{e^{+}e^{-}}} 
\newcommand{\pip}          {\ensuremath{\pi^{+}}\xspace}
\newcommand{\pim}          {\ensuremath{\pi^{-}}\xspace}
\newcommand{\kap}          {\ensuremath{\rm{K}^{+}}\xspace}
\newcommand{\kam}          {\ensuremath{\rm{K}^{-}}\xspace}
\newcommand{\pbar}         {\ensuremath{\rm\overline{p}}\xspace}
\newcommand{\kzero}        {\ensuremath{{\rm K}^{0}_{\rm{S}}}\xspace}
\newcommand{\lmb}          {\ensuremath{\Lambda}\xspace}
\newcommand{\almb}         {\ensuremath{\overline{\Lambda}}\xspace}
\newcommand{\Om}           {\ensuremath{\Omega^-}\xspace}
\newcommand{\Mo}           {\ensuremath{\overline{\Omega}^+}\xspace}
\newcommand{\X}            {\ensuremath{\Xi^-}\xspace}
\newcommand{\Ix}           {\ensuremath{\overline{\Xi}^+}\xspace}
\newcommand{\Xis}          {\ensuremath{\Xi^{\pm}}\xspace}
\newcommand{\Oms}          {\ensuremath{\Omega^{\pm}}\xspace}
\newcommand{\degree}       {\ensuremath{^{\rm o}}\xspace}

\newcommand{\XiNosign}           {\ensuremath{\Xi}\xspace}
\newcommand{\OmNosign}           {\ensuremath{\Omega}\xspace}
\newcommand{\zrange}[1]    {\mbox{$\left | z \right |<{\rm #1}$}}
\newcommand{\avdndetaplot} {\ensuremath{{\langle\dndeta\rangle}_{\etarange{0.5}}}\xspace}
\newcommand{\Xzero}            {\ensuremath{\Xi^0}\xspace}
\newcommand{\VZEROM}        {\rm{V0M}\xspace}
\newcommand{\dndetaNew}       {\ensuremath{{\langle\dndeta\rangle}_{\etarange{0.5}}}~(V0M~class)\xspace}
\newcommand{\piNosign}          {\ensuremath{\pi}\xspace}
\newcommand{\nch}       {\ensuremath{\mathrm N_\mathrm{ch}}\xspace}
\newcommand{\chindf}       {\ensuremath{\chi^{2}/NDF}\xspace}
\newcommand{\ns}       {\ensuremath{n_{\rm S}}\xspace}
\newcommand{\ynp}[2]  {\ensuremath{\rm \langle Y_{#1#2} \rangle}\xspace}
\newcommand{\dels}       {\ensuremath{{\rm \Delta S}}\xspace}

\newcommand{\pem}       {PYTHIA~8 Monash 2013\xspace}
\newcommand{\pecrr}     {PYTHIA~8 QCD-CR + Ropes\xspace}
\newcommand{\eplhc}     {EPOS~LHC\xspace}

\begin{titlepage}
\PHyear{2025}       
\PHnumber{257}      
\PHdate{05 November}  

\title{Strangeness enhancement at its extremes: multiple (multi-)strange hadron production in \pp collisions at $\mathbf{\sqrt{\textit{s}} = 5.02}$ \TeV}
\ShortTitle{Multiple (multi-)strange hadron production in \pp at \s = 5.02 \TeV}   

\Collaboration{ALICE Collaboration\thanks{See Appendix~\ref{app:collab} for the list of collaboration members}}
\ShortAuthor{ALICE Collaboration} 

\begin{abstract}
The probability to observe a specific number of strange and multi-strange hadrons (\ns), denoted as \pns, is measured by ALICE at midrapidity (\yrange 0.5) in \five proton-proton (\pp) collisions, dividing events into several multiplicity-density classes. Exploiting, for the first time, a technique based on counting the number of strange-particle candidates event-by-event, this measurement allows one to extend the study of strangeness production beyond the mean of the distribution. This constitutes a new test bench for production mechanisms, probing events with a large imbalance between strange and non-strange content. The analysis of a large-statistics data sample makes it possible to extract \pns up to a maximum \ns of 7 for \kzero, 5 for \lmb and \almb, 4 for \X and \Ix, and 2 for \Om and \Mo. From this, the probability of producing strange hadron multiplets per event is calculated, thereby enabling the extension of the study of strangeness enhancement to extreme situations where several strange quarks hadronize in a single event at midrapidity. Moreover, comparing hadron combinations with different $\it{u}$ and $\it{d}$ quark compositions and equal overall $s$ quark content, the contribution to the enhancement pattern coming from non-strangeness related mechanisms is isolated. The results are compared with state-of-the-art phenomenological models implemented in commonly used Monte Carlo event generators, including PYTHIA 8 Monash 2013, PYTHIA 8 with QCD-based Color Reconnection and Rope Hadronization (QCD-CR + Ropes), and EPOS LHC, which incorporates both partonic interactions and hydrodynamic evolution. These comparisons show that the new approach dramatically enhances the sensitivity to the different underlying physics mechanisms modeled by each generator.

\end{abstract}

\end{titlepage}

\setcounter{page}{2} 


\section{Introduction}

Strange-hadron production measurements hold a central role in the study of light-flavor hadron formation in hadronic collisions and, in general, in the phenomenological modelling of the non-perturbative quantum chromo-dynamical processes involved in the hadronization phase of extended colored systems.

An enhanced production rate of (multi-)strange particles in heavy-ion collisions with respect to hadronic collisions was originally proposed as a signature of the formation of the thermalized partonic medium called quark--gluon plasma (QGP) ~\cite{Rafelski:1980,Rafelski:1982}.
The strangeness-enhancement (SE) phenomenon was first studied at the BNL Alternating Gradient Synchrotron (AGS)~\cite{PhysRevLett.64.847,PhysRevLett.91.202301}, observed experimentally at the CERN Super Proton Synchrotron (SPS) ~\cite{Andersen:1999, Afanasiev:2002, Antinori:2010} and then confirmed at higher center-of-mass collision energies at the BNL Relativistic Heavy Ion Collider (RHIC) ~\cite{starSE} and at the CERN Large Hadron Collider (LHC) ~\cite{PbPb2,ANPbPb}.
Nonetheless, the origin of this phenomenon is still uncertain. Statistical hadronization models (SHM) have proven to be effective in reproducing strangeness production in heavy-ion collisions ~\cite{ANDRONIC2009142, WHEATON200984, VOVCHENKO2019295, FLOR2021}, 
explaining the modified probability for strange-hadron production in small collision systems with either the imposition of local flavor and charge conservation laws~\cite{Tounsi:2001ck}, or with an incomplete equilibration of the strange quantum number ~\cite{Becattini:06}.
From the experimental perspective, in the last decade the ALICE Collaboration has released a comprehensive set of measurements of strange hadron-to-pion yield ratios as a function of the average charged-particle multiplicity density at midrapidity (\multdensmid) in different collision systems and for different center-of-mass energies (\pp ~\cite{NP, pp7, pp13,ANSilvia}, \pPb ~\cite{multdeppPb5} and \PbPb ~\cite{PbPb2,ANPbPb}). The results showcase an enhancement pattern proportional to the strangeness content of the hadron and independent of the collision system or center-of-mass energy. In particular, the observation of SE in high-multiplicity \pp collisions challenges the idea of pure in-vacuum hadronization in hadron-hadron interactions~\cite{FischerSjostrand}. In addition to this, several phenomena traditionally attributed to the creation of a hydrodynamically-expanding partonic medium in A--A collisions have been observed in high-multiplicity \pp and \pPb interactions ~\cite{ridge1, ridge2, ridgecms,flowcms1, flowcms2, flowatlas, alicecollaboration2024observationpartonicflowprotonproton}.

Following all these observations in smaller systems, several phenomenological models implemented in commonly used Monte Carlo (MC) event generators ~\cite{pythiainfo,herwigmanual} have been extended to cope with particle production in high parton density environments ~\cite{PythiaCRbaryon,pythiaRopes,herwigstrangeness}.
In particular, PYTHIA~\cite{pythiainfo} is a general purpose MC generator that incorporates perturbative quantum-chromodynamics (pQCD) processes, hadronization through color string fragmentation and Multi Parton Interactions (MPI). The \pem tune also implements a basic version of the color-reconnection (CR) mechanism ~\cite{PythiaCR}. An advanced CR implementation of QCD-allowed 3-leg junctions ~\cite{PythiaCRbaryon} and a modified string-tension model in dense QCD environments are features included in the \pecrr version of the model~\cite{pythiaRopes}, with the purpose of describing the aforementioned experimental results. Other theoretical approaches implementing the interplay between a thermalized medium and in-vacuum hadronization, such as EPOS ~\cite{epos4general} and DCCI (Dynamical Core-Corona Initialization)~\cite{DCCIunif,DCCIcoco}, have been developed to describe particle production in all collision systems. 
Specifically, EPOS~\cite{epos} is based on the Gribov-Regge theory at parton level ~\cite{EPOSgrtheory}. Collective effects and modified hadronization in high-multiplicity environments are obtained by means of the ``core-corona'' mechanism ~\cite{epos4cc}, which includes two different regimes: a high-density core, implementing statistical hadronization and parametrized hydrodynamic expansion, and a low-density corona which features string fragmentation. SHM fits have been also extended to the lowest multiplicities which characterize \pp interactions~\cite{FISTsmallsys,Cleymanssmallsys,FLOR2022}.

The qualitative agreement of the revised models in reproducing SE patterns, though showing tensions when a quantitative comparison is performed, triggered additional experimental activity targeted at defining the connections between strange-particle production and local or global characteristics of the events under study in \pp collisions. In this respect, the connection of strangeness production with the presence of jets ~\cite{lk0sjets,2024}, the classification of events according to their isotropy in the final-state charged particle distribution ~\cite{ALICEspherocity}, and the tentative connection of SE to global event properties such as effective energy ~\cite{2025} are all to be considered as relevant follow-ups.  
Moreover, two-particle correlation techniques have been exploited to study strange quantum number conservation in small and large collision systems~\cite{2024BalanceF,PoissMario2025}.

In all previous yield measurements, strange-particle abundances are extracted by counting their total sum in a given data set -- dividing the sample according to topological or dynamical properties of the events or of the strange-particle candidates -- and then, after acceptance and efficiency corrections, normalizing to the total number of analyzed inelastic events in the sample. 
In this sense, these measurements account for the average of the strange-particle-multiplicity distribution underlying the physical production mechanism for a specific particle species. In this article, an extension of these results is reported: for the first time, the probability to observe a specific number of strange particles (\ns) per event, i.e. the strange-particle-multiplicity distribution \pns, is measured for \kzero, \lmb, \almb, \X, \Ix, \Om and \Mo in \pp collisions at \s = 5.02~TeV. This is performed in different charged-particle multiplicity-density classes, exploiting an analysis technique based on counting the number of strange particles event-by-event. 
This method, for the first time applied to strange particle yield measurement, allows the extraction of the probability to observe, e.g., events with a very limited overall charged-particle multiplicity density but featuring the presence of several cascades and, conversely, high-multiplicity events with a very limited number (or absence) of a single-strange \kzero mesons.
Additionally, the measurement of \pns allows the calculation of the average production yield of multiplets of a given strange particle. It is then possible to explore SE at its extremes, not limited to the ratio of \OmNosign/\piNosign, but extending to the presence of six $s$ quarks in \OmNosign doublets or \XiNosign triplets. At the same time, it is now possible to highlight the non-strangeness-related component in the SE pattern, by studying the ratio of \lmb and \kzero multiplets versus multiplicity and, in general, to explore the relative hadronization probability into different final-state strange hadron multiplets in events characterized by the presence of a given number of $s$ quarks created in the collision.

The paper is organized as follows. In Section~\ref{Expsetup}, the experimental setup is briefly discussed, while details on the analysis are provided in Sec.~\ref{analysis}. Results are presented, discussed and compared to phenomenological models in Sec. ~\ref{results}, and conclusions are summarized in Sec.~\ref{conclusion}.

\section{Experimental setup} \label{Expsetup}

ALICE (A Large Ion Collider Experiment) is the general-purpose heavy-ion experiment at the LHC that is optimized to perform tracking and particle identification down to as low as \pt $\sim$ 50 \MeV.
A detailed description of the ALICE detector and its performance during the LHC Run 1 and 2 campaigns can be found in Refs.~\cite{ALICEjinst, ALICEperf}, while in this section, the main sub-detectors used for the measurement presented in this article are briefly described. 

Tracking at central pseudorapidity (\etarange 0.9) is performed employing the Inner Tracking System (\ITS) and the Time Projection Chamber (\TPC), located inside a solenoidal magnet providing a 0.5 T magnetic field. 
The \ITS ~\cite{ALICEits} is the detector closest to the beam pipe and is composed of six cylindrical layers of silicon sensors located at radii between 3.9 and 43.0 cm from the beam line. Due to the high track density in central \PbPb collisions, the two innermost layers consist of silicon pixel detectors (SPD), which guarantee high granularity, while the two middle and two outer layers are formed by silicon drift (SDD) and micro-strip (SSD) detectors, respectively, and can be used for particle identification (PID) at very low momentum.
The \TPC ~\cite{ALICEtpc} is the main tracking detector, featuring a large cylindrical gaseous drift chamber (two drift regions divided by a central electrode) covering the full azimuth and radii between 78.8 and 248.9 cm from the beam line. Its endcaps consist of multi-wire proportional chambers with pad readout, providing up to 159 tracking points. It also provides PID information via the measurement of the specific ionization energy loss (\dEdx) in the detector gas.
PID is complemented by the Time-Of-Flight~(TOF) system ~\cite{ALICEtof}, a large array of multigap resistive plate chambers at a radial distance of 3.7 m from the beam line. Thanks to its very good time resolution of less than 60 ps, the \TOF discriminates among different particle species through a measurement of their time-of-flight.

The ALICE apparatus also includes several forward/backward pseudorapidity detectors. Among those, the \VZERO ~\cite{ALICEv0} are two sets of plastic scintillator hodoscopes located around the beam pipe on both sides of the interaction point, covering ${\rm 2.8 < \eta < 5.1}$ (\VZEROA section) and ${\rm -3.7 < \eta < -1.7}$ (\VZEROC section). The \VZERO provides the main inelastic trigger in \pp collisions by performing the logical AND of the digital signals from the A and C sides and, at the same time, it is used to classify events according to their multiplicity, by evaluating the sum of the analog signal amplitudes from the A and C sides (V0M multiplicity estimator, see Sec.~\ref{EvSelAndMultCat}).

\section{Analysis details} \label{analysis}

The analysis steps are described here, from event and charged-particle multiplicity categorization, to strange-particle candidate selection, raw multiplicity distribution building, corrections and systematic uncertainty evaluation.
The analysis procedure, though being very different from the one reported in Refs.~\cite{NP,pp7,pp13,ANSilvia}, exploits the experience documented therein in terms of event, multiplicity, and candidate selection. 
Data collected in 2017 during the LHC Run~2 operations in \pp collisions at \five are analyzed ~\cite{LHC2017}. This data sample features large statistics and very stable detector conditions, given by the rather limited time span over which all data were acquired (less than one month). Accelerator conditions were tuned to cope with the ALICE experiment requirements, featuring proton bunches sparsely populating the full orbit and a beam focusing strategy complying with the ALICE rate capability ($\beta^{*} = 10$ m). All this makes the analyzed sample the most suitable for this work.

\subsection{Event selection and multiplicity categorization}\label{EvSelAndMultCat}

A signal from both \VZEROA and \VZEROC in coincidence with the arrival of a proton bunch from each LHC direction defines the Minimum Bias (MB) trigger. A total of 1.1 $\cdot 10^{9}$ MB events, corresponding to an integrated luminosity $\mathcal{L}_{\rm int} = 2.3~{\rm nb^{-1}}$, are considered. Only events with a primary vertex (PV) positioned within \zrange{10~cm} around the nominal IP are selected for the analysis, where the coordinate $z$ is measured along the beam direction. The probability of more than one interaction occurring in one bunch crossing is smaller than $2\cdot10^{-3}$ throughout the whole data-taking period, and this residual contamination from in-bunch pile-up is reduced to a negligible level offline by excluding events with multiple vertices reconstructed with the SPD. Contamination from beam-gas interactions is removed offline by using the timing information from the \VZERO and by exploiting the correlation between the number of hits and tracks detected by the standalone \SPD ~\cite{ALICEperf}. 
The results reported here correspond to the event class \inelgz, including all inelastic events with at least one charged track in \etarange{1}, corresponding to about 75$\%$ of the total inelastic cross section. 

\begin{table}[h] 
   \caption{List of V0M multiplicity event classes (first column), percentile intervals (second column) and corresponding charged-particle pseudorapidity densities (\etarange{0.5}) in pp collisions at \five ~\cite{multEtrig5} used for the different particles under study (\kzero and \lmb in third, \XiNosign and \OmNosign in fourth column). Class names and percentile boundaries are chosen to match those defined in Ref.~\cite{ANSilvia}.}
   \centering
   \begin{tabular}{|c|c|c|c|} 
   \hline
   \multirow{2}{*}{V0M class}   &  \multirow{2}{*}{Multiplicity (\%)}   & \multicolumn{2}{c|}{ \dndetaNew }      \\
   \cline{3-4}
                            &                                       & \kzero, \lmb      & \XiNosign, \OmNosign              \\
   \hline\hline     
   $\text{I+II+III}$        & $0-0.9152$                           & $18.50 \pm 0.17$  & $18.50 \pm 0.17$                  \\
   \hline
   $\text{IV}$              & $0.9152-4.5770$                       & $14.51 \pm 0.18$  & $14.51 \pm 0.18$                  \\
   \hline
   $\text{V}$               & $4.5770-9.1560$                       & $11.93 \pm 0.11$  & \multirow{2}{*}{$11.24 \pm 0.13$} \\
   \cline{1-3}
   $\text{VI}$              & $9.1560-13.740$                       & \multirow{2}{*}{$9.70 \pm 0.13$} &                    \\
   \cline{1-2}
   \cline{4-4}
   $\text{VII}$             & $13.740-18.320$                       &                   & \multirow{2}{*}{$8.38 \pm 0.12$}  \\
   \cline{1-3}
   $\text{VIII}$            & $18.320-27.510$                       & $7.76 \pm 0.11$   &                                   \\
   \hline
   $\text{IX}$              & $27.510-36.760$                       & $6.34 \pm 0.08$   & \multirow{2}{*}{$5.78 \pm 0.08$}  \\
   \cline{1-3}
   $\text{X}$               & $36.760-46.110$                       & $5.22 \pm 0.07$   &                                   \\
   \hline
   $\text{XI}$              & $46.110-65.450$                       & $3.94 \pm 0.06$   & $3.94 \pm 0.06$                   \\
   \hline
   $\text{XII}$             & $65.450-100$                          & $2.42 \pm 0.04$   & $2.42 \pm 0.04$                   \\
   \hline
   \hline
   $\text{\inelgz}$         & $0-100$                               & \multicolumn{2}{c|}{$5.48\pm 0.07$}                   \\ 
   \hline
   \end{tabular}
   \label{tab:V0Mclass}
\end{table}

Charged-particle multiplicity density at central pseudorapidity (\etarange 0.5) is one of the main ingredients of this study. It is selected (as detailed in Ref.~\cite{multEtrig5}) by defining percentile classes in the V0M amplitude distribution (V0M class), which is proportional to the overall multiplicity in the event. This forward rapidity estimator was previously shown to be unbiased in terms of charged-to-neutral particle ratios at midrapidity~\cite{pp7} and more connected to the global event activity, rather than to local midrapidity fluctuations in particle production~\cite{2025}. For each V0M class, the average of the corresponding midrapidity multiplicity distribution is computed, fully corrected for acceptance, tracking and vertexing efficiency as well as for contamination from secondary particles. This average is denoted as \multdensmid(V0M class). Table~\ref{tab:V0Mclass} reports \multdensmid(V0M class) for all the multiplicity classes used for this study. Each V0M class corresponds to a broad midrapidity multiplicity distribution, with a relative root mean square (RMS) ranging from 70\% to 30\% from the lowest to the highest multiplicity classes reported in Tab.~\ref{tab:V0Mclass}. 
More details on the implications of multiplicity variability on this specific study are given in Sec.~\ref{par:SICPMB}, where multiplicity fluctuations and the strangeness-induced charged-particle multiplicity bias are discussed.

\subsection{Candidate selection and raw multiplicity distribution engineering}

$V^0$ and cascade candidates are reconstructed at midrapidity (\yrange 0.5) through the identification of their weak-decay topology ~\cite{PDG}: 
\begin{align*}
    & \kzero \rightarrow \pip + \pim                    && \textsc{B.R.} = (69.20 \pm 0.05)\% \\
    & \lmb (\almb) \rightarrow p (\overline{p}) + \pim (\pip) && \textsc{B.R.} = (63.9 \pm 0.5)\% \\
    & \X (\Ix) \rightarrow \lmb (\almb) + \pim (\pip)   && \textsc{B.R.} = (99.887 \pm 0.035)\% \\
    & \Om (\Mo) \rightarrow \lmb (\almb) + \kam (\kap)  && \textsc{B.R.} = (67.8 \pm 0.7)\%
\end{align*}

$V^0$ daughter tracks are first selected and then, for the cascade analysis, an additional pion or kaon (bachelor track) is associated to the topology. 
The reconstructed tracks are selected in the pseudorapidity region \etarange{0.8} to maximize the \TPC acceptance throughout their full length. Moreover, they are required to fulfill a set of quality criteria, including a minimum number of TPC crossed rows ($N_{x-rows}$) and a minimum ratio between $N_{x-rows}$ and the number of \TPC findable clusters ($N_{findable}$).
The PID is performed by means of the specific energy loss (\dEdx) in the TPC gas volume, selecting a maximum distance in resolution units ($\sigma_{\dEdx}$) from the expected average determined using a Bethe–Bloch parametrization. 
Several topological selections are applied to the reconstructed weak-decay geometry in order to ensure that the candidates are genuine strange particles and to reduce the combinatorial background. 
The distance of closest approach (DCA) of daughter tracks to the PV is required to be larger than a given threshold, to exclude primary particles in the building of the weak decay. Similarly, in the cascade analysis the reconstructed $V^0$ candidate is required to have a large DCA. On the contrary, small DCA is required between the $V^0$ daughter tracks (DCA $V^0$ daughters), and between the $V^0$ and the corresponding bachelor track in the cascade reconstruction (DCA cascade daughters).
A minimum distance in the transverse direction between the PV and the secondary or tertiary decay vertex (transverse decay radius, $R_{xy}$) is imposed to avoid large contamination from primary particles, while a maximum proper lifetime is set to avoid candidates that decay too far from the PV.
The cosine of the angle between the candidate momentum and the vector connecting the primary and secondary (or tertiary) vertices (${\rm cos\theta_{PA}}$) is a very selective variable which is set to be larger than a minimum value to avoid decays not pointing to the right vertex.
To reject residual out-of-bunch (OOB) pileup, it is required that at least one of the decay daughters has either a hit in the TOF detector or is reconstructed by both ITS and TPC.
All selections, optimized based on detailed studies done on MC simulations, are extensively described in Ref.~\cite{topological} and the values are summarized in Tab.~\ref{tab:cut}.

\begin{table}[htbp]
 \centering
  \caption{List of selection criteria applied to $V^0$ and cascade candidates.}
    \begin{tabular}{|l|c|c|} 
    \hline
    Daughter track selection            &  \kzero (\lmb)        &  \XiNosign (\OmNosign)    \\ \hline\hline
    $\text{pseudorapidity}$             &  \etarange{0.8}       &  \etarange{0.8}           \\
    $\text{$N_{x-rows}$}$               &  $>69(70)$            &  $>69(67)$                \\
    $\text{$N_{x-rows}/N_{findable}$}$    &  $\geq0.8$            &  $\geq0.8$                \\
    $\text{TPC \dEdx}$                  &  $<5(4)\sigma_{dE/dx}$ &  $<5(5.5)\sigma_{dE/dx}$ \\
    $\text{$N_{trk}$ with (TOF $||$ ITS)}$   &   $\geq1$             &  $\geq1$                  \\ \hline \hline
    Candidate selection                 &   \kzero (\lmb)       &  \XiNosign (\OmNosign)    \\ \hline\hline
    $\text{rapidity}$                   &   $\yrange 0.5$       &   $\yrange 0.5$           \\
    $\text{DCA $V^0$ daughter to PV}$     &   $>0.035(0.06)$ cm    &     $>0.11 $cm             \\
    $\text{DCA $V^0$ daughters}$          &   $<0.5(0.6)\sigma$   &    $<0.8(0.7)\sigma$      \\
    $\text{DCA $V^0$ to PV}$    &   $-$   &    $>0.04$ cm\\
    $\text{DCA bachelor to PV}$    &   $-$   &    $>0.03(0.02)$ cm \\
    $\text{DCA cascade daughters}$    &   $-$   &    $<0.8(0.7)$ cm  \\
    $\text{$R_{xy} V^0$}$   &   $>0.35(0.5)$ cm   &     $>0.6$ cm   \\
    $\text{$R_{xy}$ cascade}$   &   $-$   &     $>0.4(0.35)$ cm   \\
    $\text{Proper Lifetime (\textit{m}L/\textit{p})}$      &   $<40(30)$ cm   &    $<17(10)$ cm  \\
    $\text{$V^0$ cos$\theta_{PA}$}$     &   $>0.99(0.996)$   &     $>0.985(0.98)$ \\
    $\text{cascade cos$\theta_{PA}$}$     &   $-$   &     $>0.985(0.98)$   \\
    $\text{$\Lambda$ invariant mass window}$        &   $-$ &     $\pm 0.008$ \GeVmass \\

    \hline
    \end{tabular}
   \label{tab:cut}   
\end{table}

\begin{figure}[htb]
    \begin{center}
    \includegraphics[width = 0.95\textwidth]{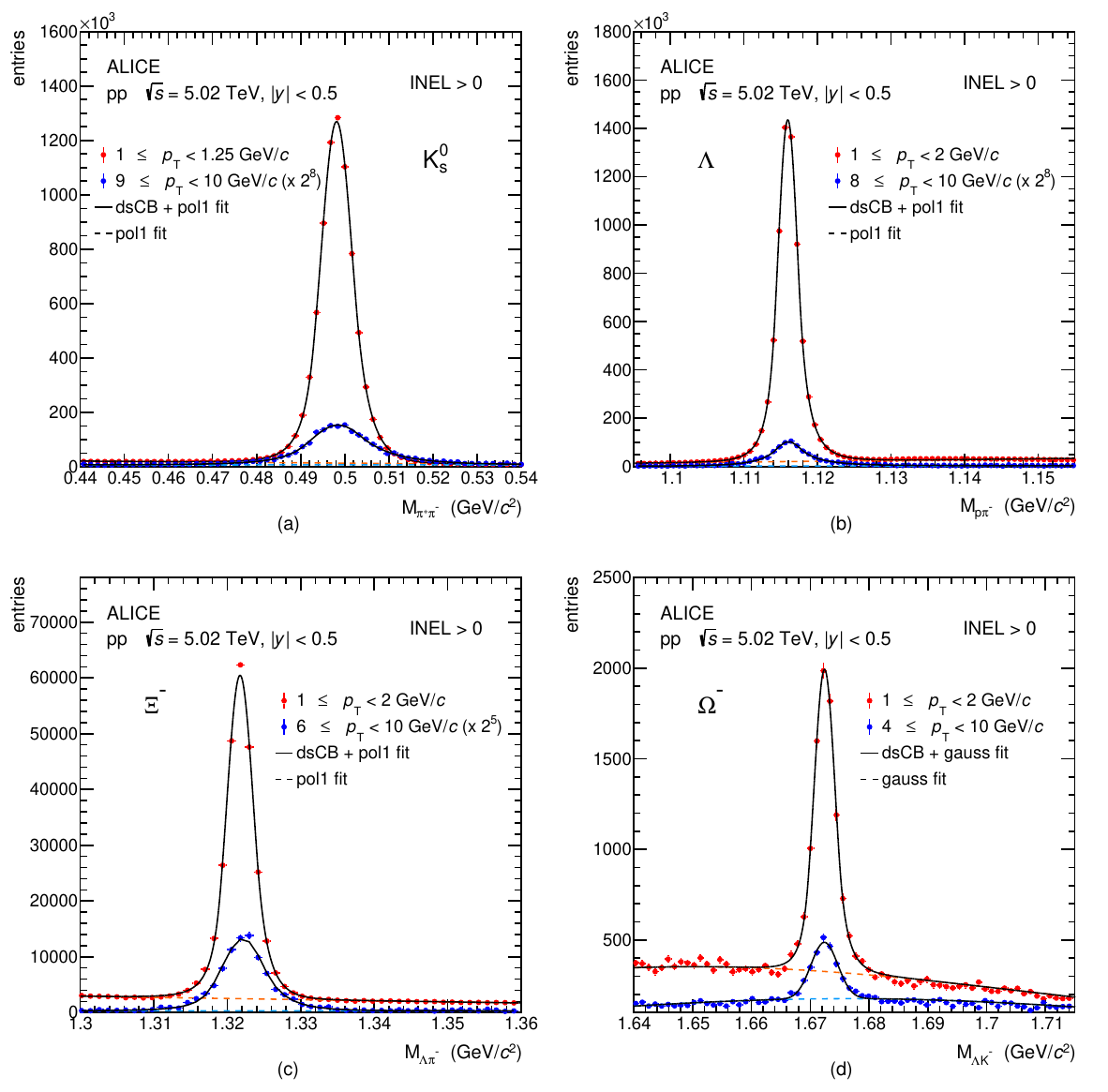}
    \end{center}
    \caption{Invariant-mass distributions for \kzero (a), \lmb (b), \X (c) and \Om (d), at low and high transverse momentum for the \inelgz event class. The dashed lines illustrate the fits used for the definition of the signal probability.}
    \label{fig:InvMassSpectra}
\end{figure}

After the application of all the aforementioned selections, the probability of a candidate to be signal or background is determined through a prior fitting procedure of the invariant-mass distribution in several \pt bins and for all multiplicity classes. A sum of a double-sided Crystal Ball (dsCB ~\cite{Gaiser:1982yw}) and a first degree polynomial (Gaussian for the \OmNosign) is used to fit the signal and the combinatorial background of the distribution, respectively. Examples of the invariant-mass distributions for strange and multi-strange particles are reported in the top and bottom panels of Fig.~\ref{fig:InvMassSpectra}, respectively, for the \inelgz event class. The distributions are reported for low and high transverse-momentum (\pt) intervals together with the fits used to determine the signal probability.  

This procedure allows the definition of weights, to be applied to each candidate, associated to its probability of being signal or background, according to its \pt, invariant mass and the multiplicity measured in the event to which it belongs:

\begin{equation}
    w_{\it S}~(\pt{\rm ,{\it m}_{inv};V0M}) = \frac{{\rm dsCB}~(\pt{\rm ,{\it m}_{inv};V0M})}{{\rm Total}~(\pt{\rm ,{\it m}_{inv};V0M})}~\{~\cdot~w_{FD}(\pt{\rm ;V0M})~\}~~ \quad , ~~ w_{\it B}~=~1-w_{\rm S}
\end{equation}
    
\begin{figure}[htb]
    \begin{center}
    \includegraphics[width = 0.49\textwidth]{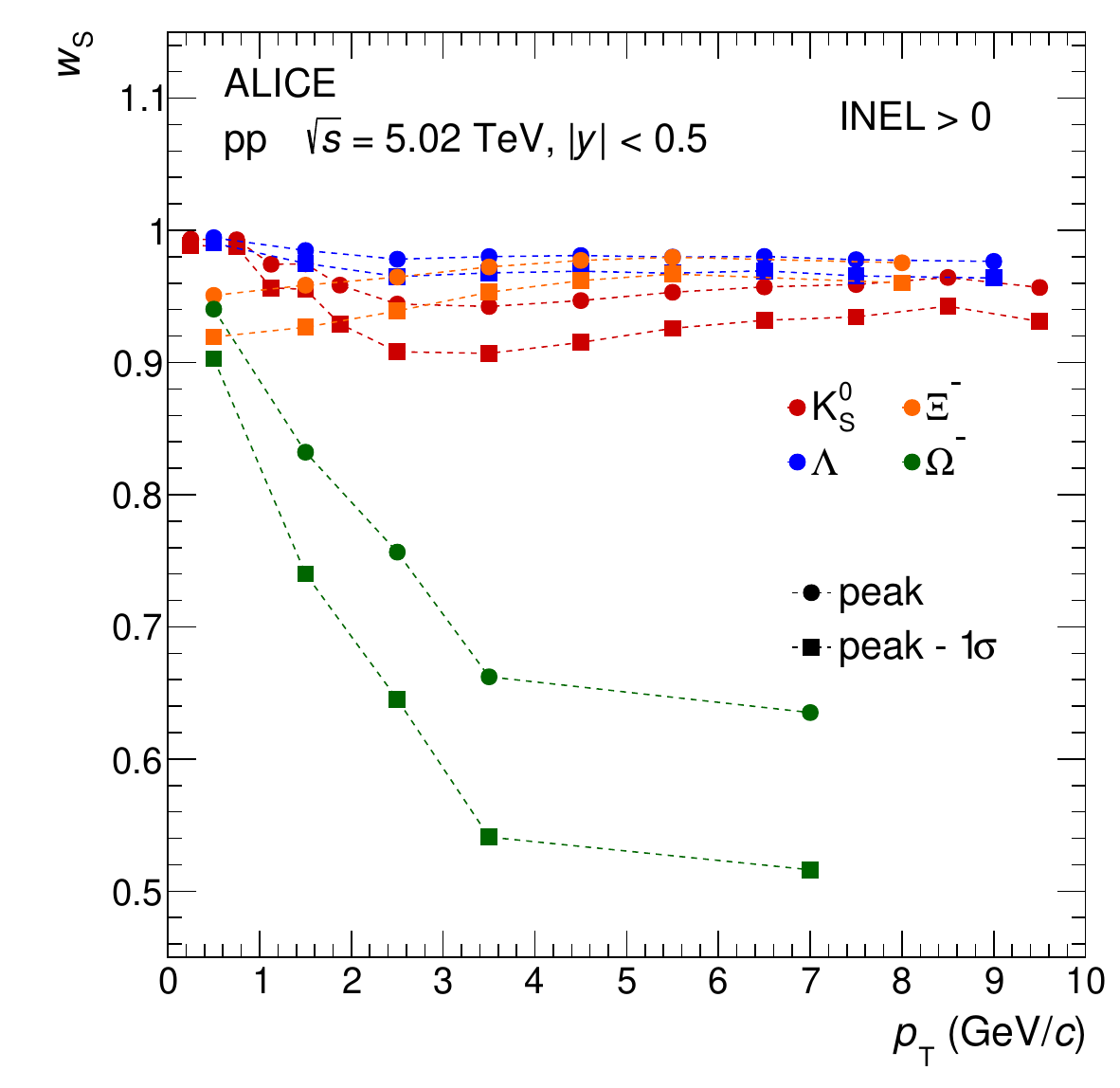}
    \hspace{0.2cm}
    \includegraphics[width = 0.48\textwidth]{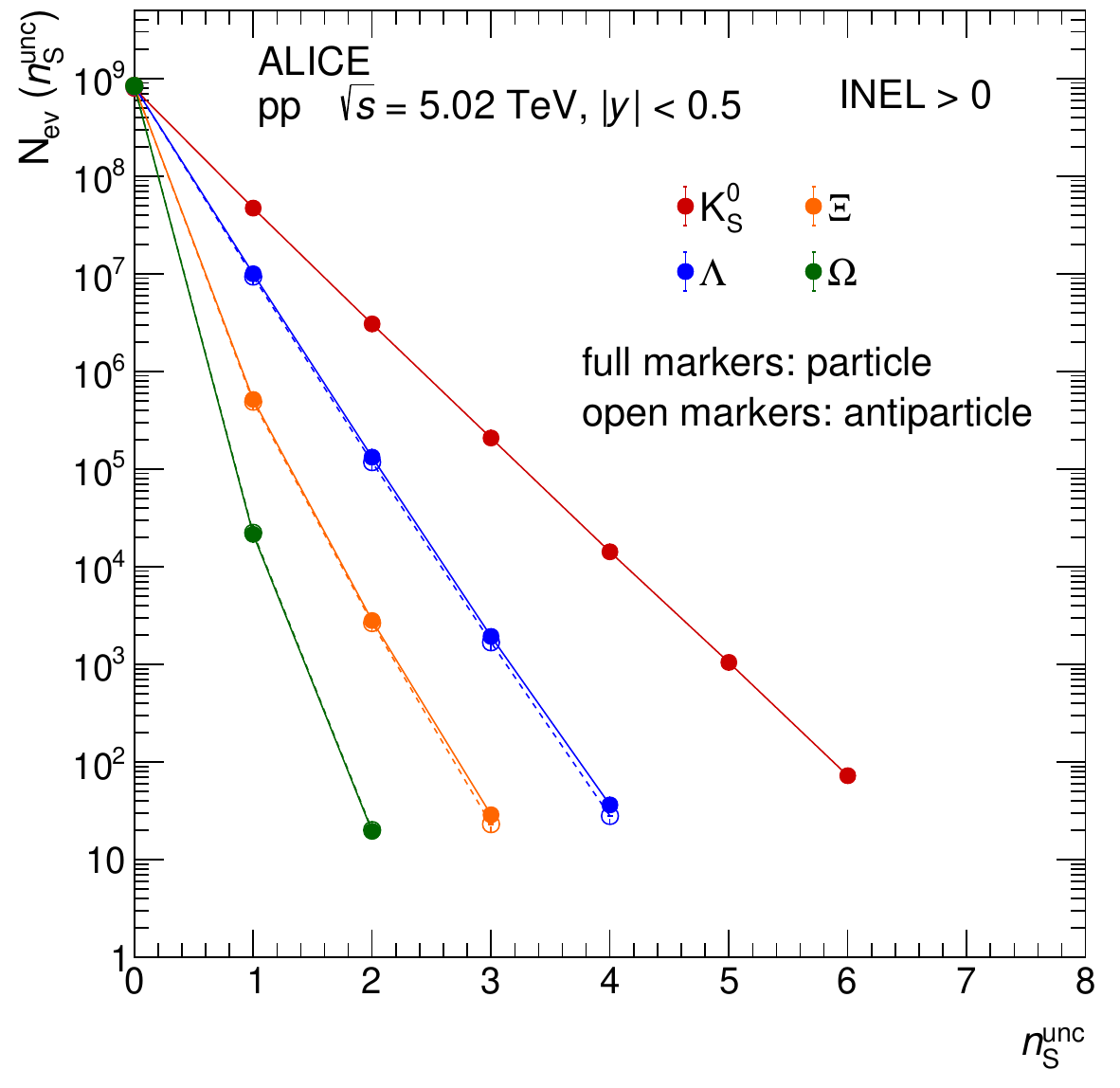}
    \end{center}
    \caption{(\textit{left}) Signal weights at the signal peak position (circles) and at 1$\sigma$ distance from the peak (square markers) for all particles under study as a function of \pt in the \inelgz event class. Dashed lines are shown to guide the eye. (\textit{right}) Uncorrected multiplicity distribution $n_{\mathrm{S}}^{unc}$ for all particles under study in the \inelgz event class, where statistical uncertainties are evaluated using the sub-samples method ~\cite{Statunc} and are smaller than the marker size. Particles correspond to full markers, antiparticles to hollow markers.}
    \label{fig:wandrec}
\end{figure}

where $w_{FD}(\pt{\rm ;V0M})$ is a factor that only applies to the \lmb(\almb) analysis, where the feed-down component coming from the decay of \X (\Ix) and \Xzero has to be taken into account. $w_{FD}$ is evaluated as the ratio between reconstructed primary and inclusive \lmb in the MC, once the \pt distributions of both \XiNosign and \lmb have been corrected at the generated level to match those measured in Ref.~\cite{ANSilvia}. Note that the \Xzero \pt shape is considered equal to the shape measured for charged \XiNosign. This procedure, different with respect to the one adopted in previous analyses and based on the determination of the feed-down matrix, relies on the previous determination of the \pt distribution for the particles under study. The comparison between the feed-down fractions determined with the two techniques shows very good agreement. For the \inelgz event class, within the measured transverse-momentum range 0 $<$ \pt $<$ 10 \GeVc, the resulting feed-down fraction shows a pronounced \pt dependence: it is about 0.11 (0.12) in the lowest \pt interval, increases to a maximum of approximately 0.23 (0.24) at \pt $\sim$ 5.5 \GeVc, and then decreases towards higher transverse momenta, reaching values of about 0.07 (0.02) in the highest \pt interval.\\
The left panel of Fig.~\ref{fig:wandrec} shows $w_S$ as a function of the candidate's \pt for all particles under study in two different regions of the invariant-mass distribution: at the peak position and at a -1$\sigma_{dsCB}$ distance from it. $w_S$ is in general very high at peak, and remains larger than 0.9 even at 1$\sigma_{dsCB}$ distance for all particles except \OmNosign, where the poorer S/B ratio translates into smaller signal probabilities. It is worth noting that $w_S$ has a smooth evolution with \pt, demonstrating that the \pt granularity chosen in this study is sufficiently large not to have $w_S$ weights rapidly changing inside the \pt bins. 

Each of the $N$ candidates in every event is then considered with its corresponding $w_{\it S}$ and the combined probability for all candidates to be signal, background and all intermediate $\frac{N!}{M!(N-M)!}$ combinations with $M$ signals and $N-M$ backgrounds are evaluated and summed.
In this way, for each event with $N$ candidates a fully reconstructed probability distribution spanning from 0 to $N$ is extracted. 
As an example, with three $V^0$ candidates in the event $i$, the reconstructed distribution $R_i$ is composed by four bins, whose contents are calculated as:
\begin{align*}
    & R_{i}[0] = w_{B}(1) \cdot w_{B}(2) \cdot w_{B}(3)\\
    & R_{i}[1] = w_{S}(1) \cdot w_{B}(2) \cdot w_{B}(3) + w_{B}(1) \cdot w_{S}(2) \cdot w_{B}(3) + w_{B}(1) \cdot w_{B}(2) \cdot w_{S}(3)\\
    & R_{i}[2] = w_{S}(1) \cdot w_{S}(2) \cdot w_{B}(3) + w_{S}(1) \cdot w_{B}(2) \cdot w_{S}(3) + w_{B}(1) \cdot w_{S}(2) \cdot w_{S}(3)\\
    & R_{i}[3] = w_{S}(1) \cdot w_{S}(2) \cdot w_{S}(3).
\end{align*}
Summing over all the analyzed events (over all $i$ in the example above) and normalizing to the total number of events, the fully uncorrected distribution is extracted. The procedure is carried out for the \inelgz event class and in every V0M multiplicity class. 
The right panel of Fig.~\ref{fig:wandrec} reports the uncorrected multiplicity distribution for all particles and antiparticles before normalization, where statistical errors have been evaluated using the sub-samples method ~\cite{Statunc}. The distribution is truncated at an $n_{\mathrm{S}}^{unc}$ which features at least 10 occurrences in the analyzed sample.\\

\subsection{Corrections: unfolding and trigger efficiency}

The correction for detector response, normally applied to estimate the acceptance and efficiency factor (\acceff), is here performed by means of a one-dimensional Bayesian unfolding procedure implemented in the package \textit{RooUnfold} ~\cite{roounf}.
A dedicated MC simulation has been performed featuring \pem events plus injected strange particles generated according to their measured \pt distribution from Ref.~\cite{ANSilvia}.
Strange hadrons simulated by PYTHIA are removed a-posteriori, as their \pt distribution is different from the one observed in real data and, hence, they would introduce a bias in the procedure.
Realistic detector conditions, as monitored during the whole data-taking period, are taken into account in the simulation by implementing particle propagation through the detector material with the GEANT~4~\cite{Brun:1993} transport code. The track reconstruction and candidate selection are identical to what is done for real events. On top of all the selection criteria previously discussed, the correspondence between the reconstructed candidate and the originally generated particle of the correct type is ensured.
Strange particles are injected with a sub-sample scheme ensuring to have similar statistics in events featuring the presence of very different strange-particle multiplicities. This results in a flat distribution of the number of events as a function of the generated strange-particle multiplicity.

\begin{figure}[htb]
    \begin{center}
    \includegraphics[width = 0.97\textwidth]{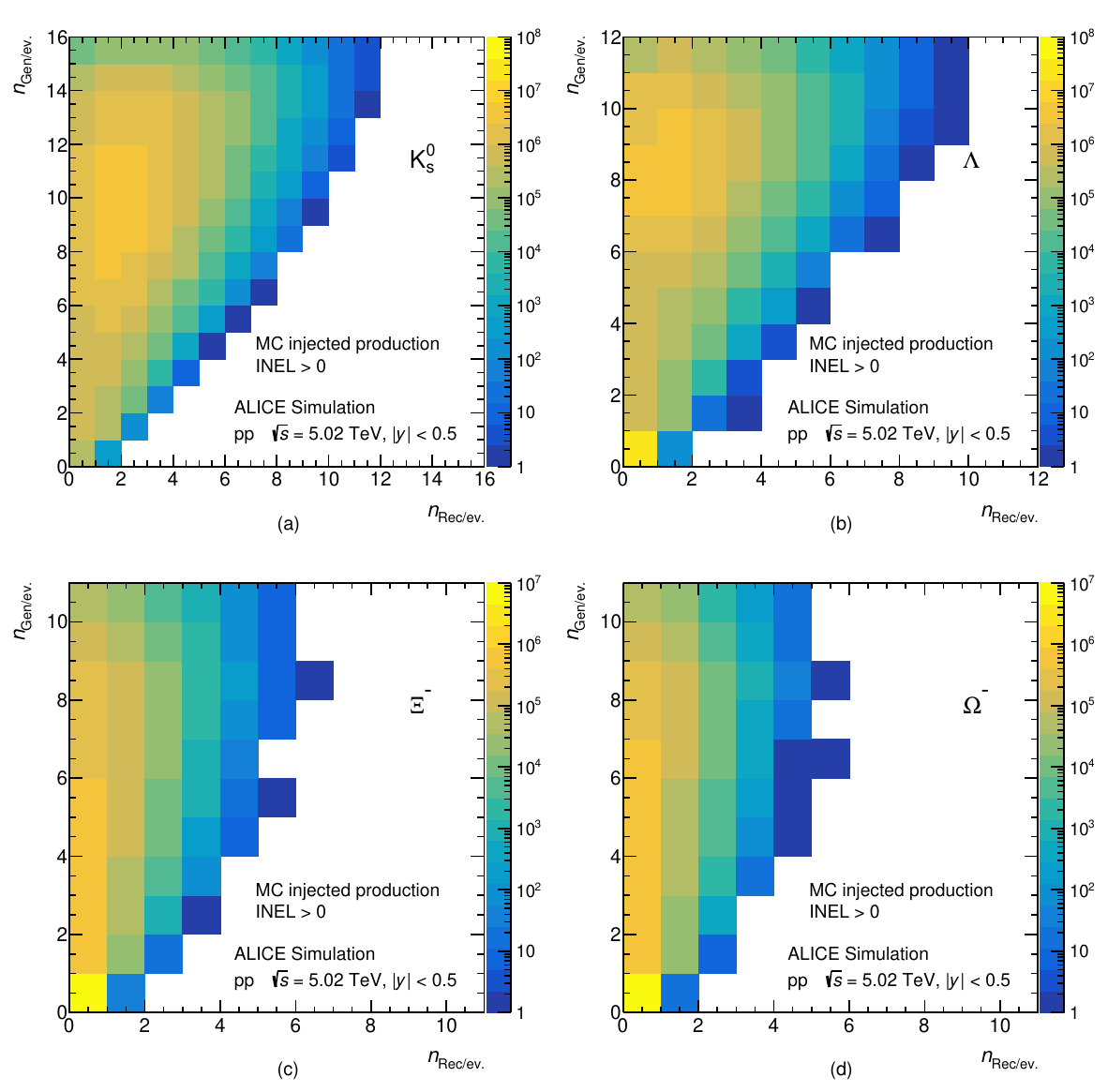}
    \end{center}
    \caption{Response matrices showing the number of generated particles per event as a function of the number of reconstructed particles per event, used for the multiplicity measurement of \kzero (a), \lmb (b), \X (c) and \Om (d) in \pp collisions at \five for the \inelgz event class.}
    \label{fig:respomtx}
\end{figure}

Figure~\ref{fig:respomtx} shows the response matrices in the \inelgz class for \kzero, \lmb, \X and \Om. Since the \pt-averaged \acceff corresponding to the implemented selections is generally lower than 10\% for all particles, the response matrix turns out to be rather limited in the number of reconstructed particles, especially for cascades. This translates into large correction factors for the high tail of the multiplicity distribution, factors that need to be accurately studied in terms of stability. The regularization parameter in the Bayesian unfolding procedure corresponds to the number of iterations needed to stabilize the result. It is observed that in the first $\sim10$ iterations the unfolded distribution changes significantly, as the starting point in the response matrix is a flat generated distribution, while the result turns out to be a rapidly-falling distribution. Generally, after 10 iterations the unfolded result fluctuates less than $\sim1\%$ if the number of iterations is increased. For this reason, after the first 10 iterations, a preliminary unfolded distribution is fitted with a negative binomial distribution (NBD ~\cite{NBDfit}), which reproduces the data well (see Sec.~\ref{results}), and a new unfolding procedure is started by weighing the generated distribution in the response matrix to correspond to the extracted preliminary NBD. In this configuration the Bayesian procedure necessitates a smaller regularization parameter to converge (typically 3–4 iterations at most), and the MC closure test verifies this procedure starting from different generated multiplicity distributions at the $\sim$ 2\% level for all particles. The evaluation of a systematic uncertainty on the prior choice for the unfolding procedure is detailed in Section~\ref{sec:systs}.

It has been previously shown~\cite{ANSilvia} that the \pt-differential \acceff does not depend on the event multiplicity for any of the particles under study in \pp collisions. Nonetheless, the evolution with multiplicity of the \pt distribution associated to strange-particle production translates, when integrating over \pt, into a different average \acceff as the \VZEROM class changes. 
For this reason, detector response in different multiplicity classes has been obtained starting from the same MC simulation and applying, to every candidate, suitable weights to arrange the \pt distribution to the measured one in the \VZEROM class under study.

The statistical uncertainty related to the particle-multiplicity distribution is estimated using the sub-sample method ~\cite{Statunc}. The initial dataset is divided into 20 sub-samples for $V^0$ and 15 for cascades, all of equal size. For each sub-sample the multiplicity distribution is obtained following the procedure previously discussed. The central point and the statistical uncertainty associated to each bin of the multiplicity distribution are obtained as the mean and the standard deviation of the 20 (15) obtained values. This procedure serves also as an important test for the stability of the whole procedure, since the distribution of the 20 (15) values turns out to be peaked at the mean value, rapidly falling symmetrically. There are no outliers observed in the distribution, which demonstrates that the unfolding procedure is stable.

Event trigger inefficiency affects the measurement in two distinct ways. 
Firstly, the normalization of the strange-particle-multiplicity distribution to the total number of analyzed events should take into account the trigger efficiency factor ($\varepsilon_{trig}$) estimated and reported in Ref.~\cite{multEtrig5}. $\varepsilon_{trig}$ is applied differentially to each bin of the multiplicity distribution, taking into account its \dndeta dependence and the multiplicity bias discussed in Sec.~\ref{par:SICPMB}.
Secondly, lost events may contain strange-particle candidates and, for this reason, a correction factor that counts the number of lost candidates ($\varepsilon_{part}$) is estimated from MC simulations. In this case, the \pt-differential factor evaluated in Ref.~\cite{ANSilvia} is averaged taking into account the correct \pt spectrum for all particles in every multiplicity class. $\varepsilon_{part}$ is applied to each bin $n$ of the multiplicity distribution elevated to the $n$-th power.

As a result, the multiplicity distribution for the strange-particle ($S$) production is obtained as:
\begin{equation}
 {\rm P}(n_{\rm S}) = \frac{\varepsilon_{trig}}{\varepsilon_{part}^{n}} \cdot \left\{ \frac{\sum_{i=1}^{N_{tot}}R_{i}[n]}{N_{tot}}\right\}_{Unfolded},
\end{equation}
where $N_{tot}$ is the total number of analyzed events.

\subsection{Systematic uncertainty evaluation} \label{sec:systs}

Several sources of systematic uncertainty are considered, and the effect on the final results is evaluated. Table~\ref{tab:systematics} reports the relative systematic uncertainty estimated in the \inelgz event class for all particles under study at the lowest (zero), intermediate and highest \ns considered. The general outcome is that the systematic uncertainty smoothly increases with \ns, as the effect of any bias contributes more as \ns gets higher: for $n_{\mathrm{S}} =$ 0 the multiplicity dependent systematic uncertainties are always smaller than 4\%, while they reach tens of percent for the highest values of \ns.

\begin{table}[htbp]
 \centering
  \caption{Systematic-uncertainty sources and corresponding values in three different \ns (zero, intermediate and highest) of the \inelgz multiplicity distribution of all strange hadrons under study. The last two sources are evaluated separately for particles and antiparticles, with only the highest value reported in the table (in parentheses if it is the one corresponding to antiparticles).}
  \resizebox{\textwidth}{!}{
    \begin{tabular}{|l|c|c|c|c|c|c|c|c|c|c|c|c|} 
    \hline
                & \multicolumn{12}{c|}{Systematic uncertainty (\%)} \\
     \cline{2-13}
     Hadron     & \multicolumn{3}{c|}{\kzero}   &   \multicolumn{3}{c|}{\lmb}   & \multicolumn{3}{c|}{\XiNosign}    & \multicolumn{3}{c|}{$\Omega$}   \\ 
     \cline{2-13}
     \ns      & 0     & 3        & 7          &   0   & 2      & 5            & 0     & 2       & 4               & 0     & 1       & 2                 \\ 
     \hline\hline
     $\text{Cut variation}$             & 0.3   & 5.6   & 21                & 0.07  & 5.7   & 38            &0      & 6.3      & 38               & 0       & 4.4   & 10\\
     $\text{Signal extraction}$         & 0.09  & 0.4   & 0.6               & 0     & 0.17   & 0.9           &0      & 0.19     & 0.23              & 0.01    & 0.04   & 0.07\\
     $\text{Unfolding procedure}$       & 1.6   & 17.8   & 22.5              & 0.4   & 16.8   & 65            &0.2    & 15.1    & 65                 & 0       & 10.7   & 24    \\
     $\text{$\varepsilon_{part}$}$      & 0.9   & 6.3   & 15.8              & 0.2   & 3.7   & 9.5           &0.04   & 7.3    & 14                & 0        & 2.95   & 5.9\\
     $\text{\pt distribution}$          & 0.4   & 11.9   & 12.5              & 3.5   & 4.9   & 45            &3      & 3.3    & 5.6               & 0        & 5.4   & 13\\ 
     $\text{OOB pile-up cut}$           & 0.09  & 8.5   & 16.5              & 0.4   & 9.8   & 26            &0.01   & 2.6    & 5                 & 0        & 0.43   & 0.85\\
     $\text{Feed-down correction}$      & -     & -   & -                 & 0.15  & 3.01   & 7.8           & -     & -    & -                 & -        & -   & - \\
     $\text{material budget}$           & 0.6   & 1.6   & 3.7               & 1.1   & 2.2   & 5.4           &1.2    & 2.3    & 4.7               & 1        & 1.1   & 2\\
     $\text{Strangeness from material}$ & 0.15  & 0.38   & 0.9               & 0.4   & 0.7   & 1.7           &0.48   & 0.9    & 1.8               & 0.25     & 0.31  & 0.44\\
     $\text{Hadronic interaction}$      & -     & -   & -                 & (0.17)& (0.31) & (0.8)         & (0.2)    & (0.36)   & (0.72)             & (0.2)    & (0.28)  & (0.36)\\
     \hline
     \hline
     $\text{Total}$                     & 2     & 23.5    & 37                & 3.7  & 21.5    & 90            & 3.3  & 17.6    & 66                & 1.1      & 14.4    & 30\\
     \hline
    \end{tabular}
    }
   \label{tab:systematics}   
\end{table}

Potential differences in the distribution of selection variables in MC and real data can impact the final result. To study this, a multi-trial procedure is followed.
For each selection variable, a window around the default value is defined by identifying two extreme values which lead to a raw signal increase or decrease of no more than 10\% with respect to the default. Afterward, 500 different combinations of selection criteria are randomly chosen inside their respective variability window and the full analysis chain is performed for every trial, from signal-extraction to raw distribution determination, unfolding and trigger-efficiency corrections.
For the \lmb and \almb analyses the feed-down weight is also re-evaluated for each trial, as the correction is applied at the reconstructed-candidate level.
For each bin of \pns, the procedure leads to 500 values which are Gaussian distributed around the default and the standard deviation of the distribution defines the systematic uncertainty. Given the complex and resource-consuming procedure, this systematic uncertainty is evaluated on the \pns in the \inelgz class and then applied to all the \pns in the different \VZEROM classes.\\
The systematic uncertainty on the signal-extraction procedure is evaluated by changing the \pt granularity in the determination of $w_S$. These weights depend on \pt (as shown in Fig.~\ref{fig:wandrec} (\textit{left})) because the selection variables have a \pt dependence and the combinatorial background evolves with the transverse momentum. The default binning in \pt is chosen to guarantee sufficient statistics in every invariant-mass distribution, so that a reliable fit can be performed. The systematic uncertainty is evaluated by slightly changing the number of \pt bins (12, 11, 11 and 9 for \kzero, \lmb, \XiNosign and \OmNosign, respectively) and following the full analysis procedure for the modified binning configuration. The difference between the new result and the default one is considered as a systematic uncertainty. The effect is sub-leading, never reaching 1\% for any \ns, demonstrating that the \pt granularity chosen in the default analysis is sufficient to guarantee a solid $w_S$ determination.\\
As previously mentioned, the stability of the unfolding procedure is extensively tested. In this case, a systematic uncertainty can originate from the choice of the prior generated multiplicity distribution in the response matrix. In the default procedure, the NBD obtained from a fit to the unfolded distribution after 10 initial iterations is used as a starting point. Two different functions for each \VZEROM class $i$ are considered, namely the preliminary NBD fit corresponding to classes $i-1$ and $i+1$. For the lowest(highest) \VZEROM class, where $i-1$($i+1$) is absent, the $i+2$($i-2$) case is considered, while in the \inelgz case the two adjacent classes V and VI are used. 
The largest difference between the result obtained with the default approach and the two alternative ones is retained as systematic uncertainty. At low \ns, the relative uncertainty is very small, while it increases significantly for higher \ns, becoming the dominant one for the highest \ns for all particles. Low-multiplicity classes are those where the effect is the highest, while the uncertainty decreases at high multiplicity.\\
Corrections due to trigger inefficiencies can also be sources of systematic uncertainty. $\varepsilon_{trig}$ is known to a very high precision ~\cite{multEtrig5} and, therefore, cannot lead to sizeable systematic uncertainty. On the contrary, $\varepsilon_{part}$ is model dependent and different MC generators are used to estimate variations on this correction factor. This procedure is the one followed in Ref.~\cite{ANSilvia}, leading to a \pt-dependent uncertainty ($\delta\varepsilon_{part}$) which can be exploited to study the effect on the strange-particle-multiplicity distribution determination. Applying two different $\varepsilon_{part}$ factors corresponding to $\varepsilon_{part}\pm\delta\varepsilon_{part}$, two new corrected \pns for each \VZEROM class are obtained and the maximum variation with respect to each default case is considered as systematic uncertainty.\\
One of the key ingredients in the analysis reported here, that allows one to perform a one-dimensional unfolding procedure, is the previous determination of the \pt distribution for all hadrons under study~\cite{ANSilvia}. Transverse-momentum spectra are known up to a given experimental precision and their shape in the un-measured low-\pt region is extrapolated, so a systematic uncertainty can arise by the use of a different \pt parametrization.
For this reason, the \inelgz \pt spectrum is interpolated at low \pt by re-scaling two functional forms associated to different \VZEROM classes (one leading to larger and one to smaller extrapolation factors) and the response matrix is reweighted with the new \pt parametrizations. The full analysis is then performed in both cases and the maximum difference with respect to the default procedure is retained as systematic uncertainty.\\
The systematic uncertainty due to the OOB pileup rejection is obtained by changing the number of daughter tracks that are required to match the (TOF$||$ITS) condition (see Table~\ref{tab:cut}) to at least two, therefore tightening the cut with respect to the default case. The difference in the final result, only significant for the highest \ns in the $V^0$ analysis, is considered as systematic uncertainty.\\
\lmb and \almb distributions are also affected by a systematic bias coming from the feed-down evaluation, given the uncertainty in the knowledge of the \XiNosign \pt distributions. The feed-down fraction is re-calculated varying the measured cascade \pt spectra inside their respective systematic uncertainties, which leads to slightly different $w_{FD}$ factors that are applied to perform signal extraction. After performing the full analysis procedure to these varied reconstructed distributions, the relative systematic uncertainty is obtained evaluating the maximum difference with respect to the default case.\\
The imperfect reproduction of the detector material budget in the MC is also taken into account. The uncertainty on the material budget determination is discussed in Ref.~\cite{ALICEperf} and amounts to $\pm 4.5\%$. In several previous analyses ~\cite{pp13, pp7,ANPbPb}, the effect of this uncertainty on the \acceff for strange-particle measurements has been explored and reported. To translate this systematic effect on the multiplicity distribution determination, the \pt-dependent uncertainty is weighted for the measured \pt distribution, thus obtaining a $\delta_{mat}$ factor which accounts for the efficiency bias associated to each reconstructed candidate. For this reason, the systematic uncertainty due to material budget determination is obtained as $(1-\delta_{mat})^{n_{\mathrm{S}}}$. \\
Finally, two minor sources of systematic uncertainty come from the limited knowledge of the cross section for strange-particle formation in the interaction of primary hadrons with the detector material and for the annihilation of strange antiparticles with the detector material. In both cases, the results reported in previous publications (~\cite{ANPbPb}) are considered. Like for the material budget uncertainty, \pt-dependent systematics are weighed for the measured \pt distribution and applied for each candidate. The relative systematic uncertainty is negligible with respect to all other sources.

\subsection{Average yield of strange-hadron multiplets}
\pns is the measurement of the probability of observing a specific number of strange particles (\ns) in the final state. From this strange-particle-multiplicity distribution, the average production yield of $n$ strange particles per event (multiplets of order $n$) can be calculated through: 
\begin{equation}
  \ynp{n}{S} = \sum_{i=n}^{n_{S}^{max}} \frac{i!}{n!(i-n)!} \cdot P(i_{S}),
  \label{eq:yields}
\end{equation}
where $n_{\mathrm{S}}^{max}$ is the highest measured bin of the multiplicity distribution for particle $S$.
The combinatorial factor in front of P($\rm i_S$) takes into account all the combinations of $\rm i_S$ strange particles contributing to the order-$n$ average production yield:
\begin{align*}
  &\ynp{1}{S} = P(1_{S}) + 2\cdot P(2_S) + 3\cdot P(3_S) + \cdots\\
  &\ynp{2}{S} = P(2_S) + 3\cdot P(3_S) + 6\cdot P(4_S) + \cdots\\
  &\ynp{3}{S} = P(3_S) + 4\cdot P(4_S) + 10\cdot P(5_S) + \cdots\\
  &\cdots
\end{align*}
\ynp{1}{S} is the average of the multiplicity distribution, hence the mean rate for the production of hadron $S$, corresponding to what was called $\langle \dndy \rangle$ or ``yield'' in previous publications~\cite{NP,pp7,pp13,ANSilvia}. As such, a comparison to previous results is possible, thus allowing to check the solidity of this new analysis technique. \ynp{n>1}{S} identifies the average production yield of multiplets (doublets, triplets, etc.) of the given particle.
Statistical and systematic uncertainty on each bin of the multiplicity distribution are considered as fully uncorrelated and propagated accordingly in Eq.~\ref{eq:yields}. 
Also in this case, uncertainties are comparable to previous results for \ynp{1}{S}, demonstrating perfect agreement.

It is worth noting that Eq.~\ref{eq:yields} would theoretically extend to infinity, as in the determination of the multiplets' probability, the whole multiplicity distribution counts. Nonetheless, as it will be clear from Sec.~\ref{results}, \pns are rapidly falling distributions, with a few orders of magnitude difference in the probability associated to contiguous bins. For this reason, one can naively expect that truncating the calculation to $n_{\mathrm{S}}^{max}$ introduces a small bias in the \ynp{n}{S} determination. As a general rule, the $n$-plet determination is only attempted up to $n = n_{\mathrm{S}}^{max}-1$ for all particles except \OmNosign, where the calculation is extended to $n = n_{\OmNosign}^{max}$. An additional source of systematic uncertainty associated to the \ynp{n}{S} determination is added by estimating the bias introduced by the sum truncation. An extrapolation is performed on the tail of the multiplicity distribution, by fitting its last three bins with an exponential function, and the sum in Eq.~\ref{eq:yields} is extended up to $n = 100$, where P($n_{\mathrm{S}} > n^{max}$) are taken from the extrapolated values. The relative difference with respect to the default approach is considered as a one-tailed systematic uncertainty (which affects the yields upwards). Table~\ref{tab:systyield} reports the relative systematic uncertainty due to the sum truncation for all average $n$-plet yields in the \inelgz event class. A similar procedure is followed for each \VZEROM class considered. As it is clear from the table, this uncertainty increases dramatically with the order $n$ of the multiplet, because the fast decrease in \pns is balanced by a larger combinatorial factor in Eq.~\ref{eq:yields}. For this reason, the calculation of \ynp{6}{\kzero} is dropped.

\begin{table}[htbp]
 \small
 \centering
  \caption{Relative systematic uncertainty related to the sum truncation of Eq.~\ref{eq:yields} in the \ynp{n}{S} determination for the \inelgz event class.}
    \begin{tabular}{|c|c|c|c|c|c|} 
    \hline
     \multirow{2}{*}{hadron}        & \multicolumn{5}{c|}{Sum truncation systematic uncertainty (\%)}                 \\ 
     \cline{2-6}
                                    & \ynp{1}{S} & \ynp{2}{S} & \ynp{3}{S} & \ynp{4}{S} & \ynp{5}{S}  \\ 
     \hline\hline
     \kzero                         & $<$1     & 1.3      & 6.6      & 23.7    & 79.0            \\
     \lmb                           & $<$1   & $<$1     & 1.5     & 6.1    & -               \\
     \XiNosign                      & $<$1     & $<$1      & $<$1     & -          & -            \\
     \OmNosign                      & $<$1      & $<$1        & -          & -          & -             \\
    \hline
    \end{tabular}
   \label{tab:systyield}   
\end{table}

\section{Results and discussion \label{results}} 

\begin{figure}[htb]
    \begin{center}
    \includegraphics[width = 0.99\textwidth]{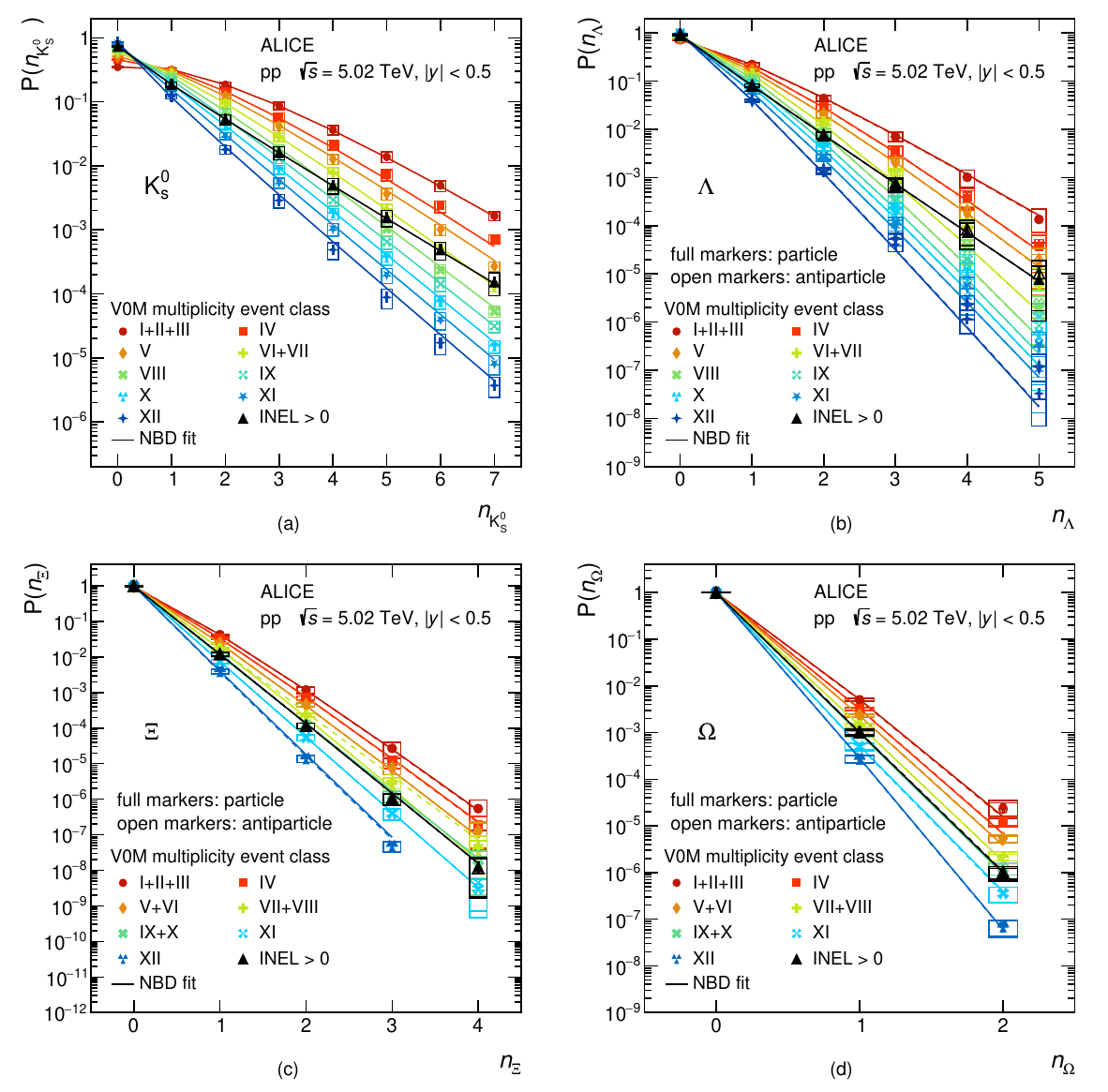}
    \end{center}
    \caption{(Multi-)strange-particle-multiplicity distributions (\pns) for \kzero (a), \lmb and \almb (b), \X and \Ix (c) and \Om and \Mo (d) for several \VZEROM multiplicity classes. Continuous (dashed) lines are the NBD ~\cite{NBDfit} fit for particles (antiparticles) to the multiplicity distributions.}
    \label{fig:Distributions}
\end{figure}

The strange-particle-multiplicity distributions \pns measured in \pp collisions at \five are reported in Fig.~\ref{fig:Distributions} for all the hadrons under study in several \VZEROM multiplicity classes. Particle and antiparticle distributions are compatible with each other within systematic uncertainties, as previously reported for the average probability ~\cite{NP,pp7,pp13}. For this reason, in the following text \lmb, \XiNosign and \OmNosign will refer to both particles and antiparticles.

With the data sample under study, \pns can be measured up to 7 \kzero, 5 \lmb, 4 \XiNosign (3 in the lowest multiplicity bin) and 2 \OmNosign particles per event. On account of this, the result represents a unique opportunity to test the connection between charged and strange particle multiplicity production throughout very extreme situations, e.g. spanning from events with 7 \kzero at low average multiplicity -- where \avdndeta $\sim 3$ -- to events with zero \kzero at high-multiplicity -- where \avdndeta $\sim 20$. 
\pns with $n > 0$ increases as a function of the multiplicity, consistently with what was previously reported for the average production~~\cite{ANSilvia}. The increase is steeper for higher $n$, as shown by the larger spacing between the curves corresponding to low- and high- multiplicity classes in the high-$n$ tail of the distribution when plotted on a logarithmic scale. 
In the same figure, the NBD ~\cite{NBDfit} fit to all multiplicity distributions is shown. The NBD discrete distribution is able to accommodate all the data, similarly to what previously observed for the charged-particle-multiplicity distribution ~\cite{Pnch09, ValentinaPnch}, with $\chi^{2}$ lower than 1 for V0s and lower than 3 for cascades.

In Fig.~\ref{fig:Distributions}, the multiplicity-integrated (\inelgz) result is also reported and it is worth noting that the shape of the \inelgz multiplicity distribution falls less steeply at high \ns with respect to the multiplicity differential distributions. This is particularly visible for \kzero and \lmb, where \pns extends to larger \ns.
This reflects a key feature, which is important for exploring the correct interpretation of the results reported here (see next section).

\subsection{The strangeness-induced charged-particle multiplicity bias} \label{par:SICPMB}
\begin{figure}[htb]
    \begin{center}
    \includegraphics[width = 0.49\textwidth]{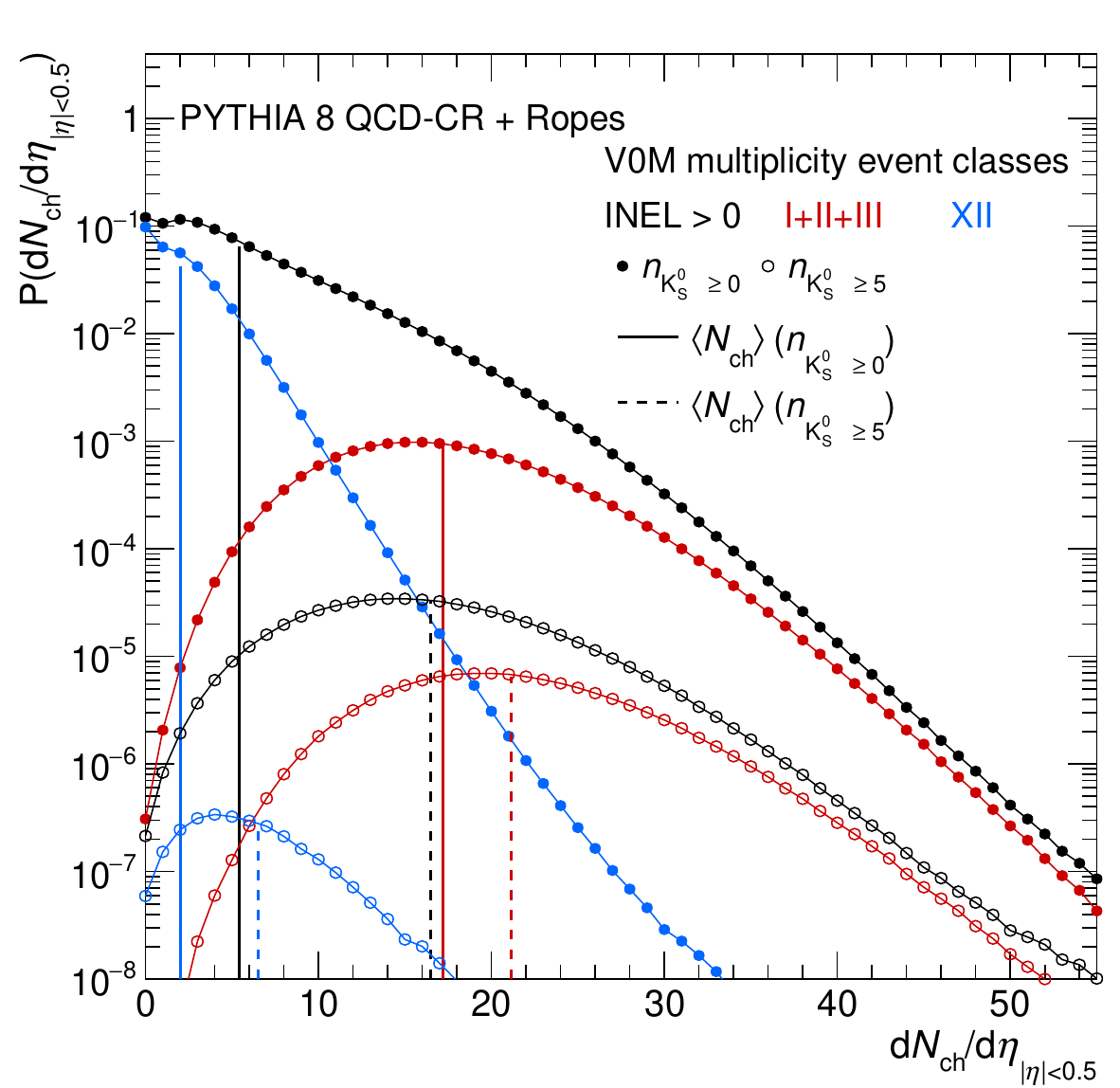}
    \includegraphics[width = 0.49\textwidth]{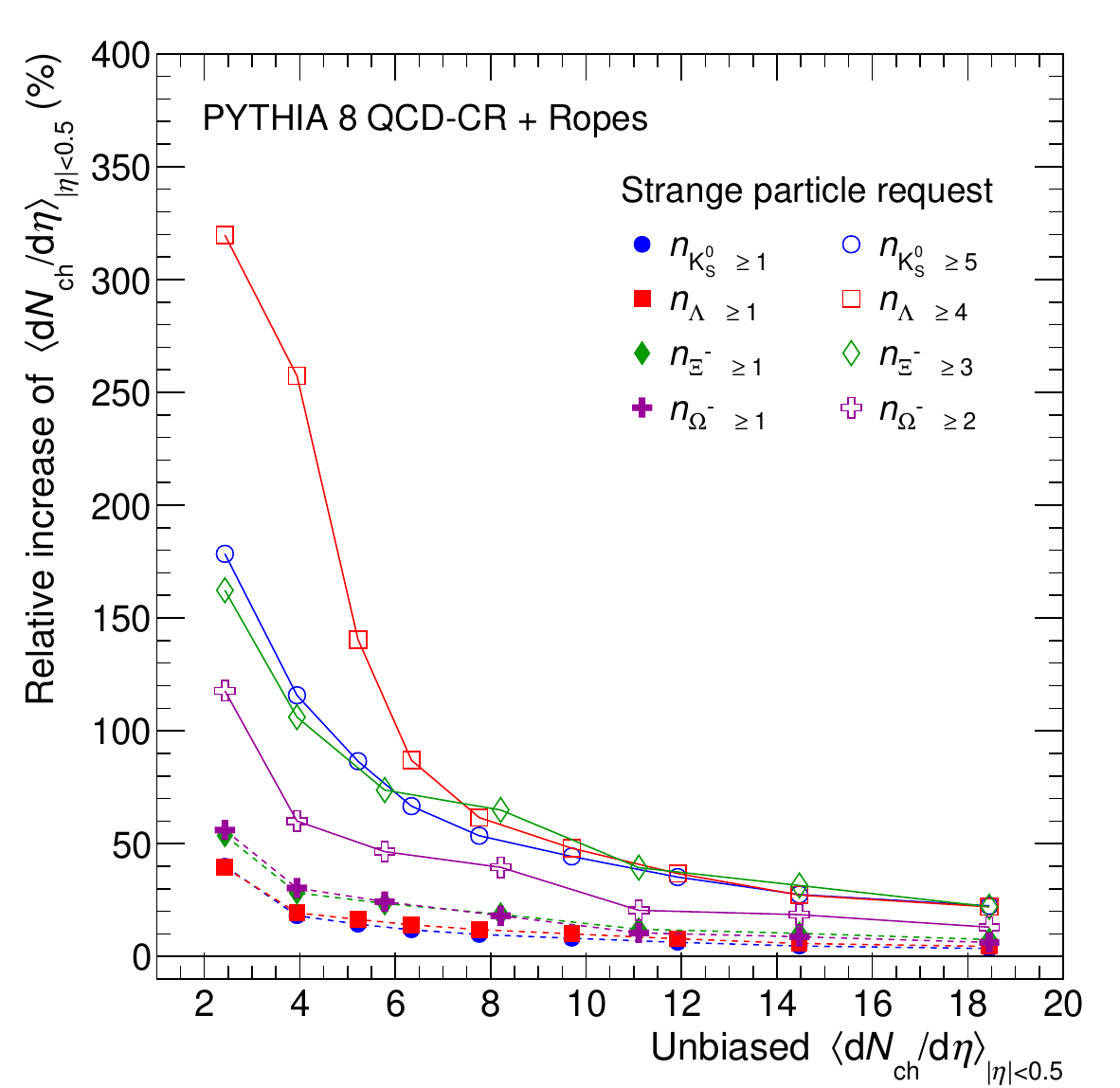}
    \end{center}
    \caption{Illustration of the strangeness-induced charged-particle multiplicity bias effect from \pecrr simulation. (\textit{left}) \dndetamid distribution in the \inelgz, \VZEROM-(I+II+III) and \VZEROM-XII event classes. Open markers are the corresponding \dndetamid resulting from the additional request of at least 5 \kzero in the event. Continuous and dashed vertical lines show the averages of the unbiased and biased multiplicity distributions, respectively. (\textit{right}) Relative shift in the \avdndeta for all the different \VZEROM event classes when requiring at least one (full markers) or more than one (open markers) strange particles in the event. Lines are shown to guide the eye.}
    \label{fig:multbias}
\end{figure}

The charged-particle-multiplicity distribution at midrapidity spans over a large range, from zero to more than 50 charged particles per event, as reported in Ref.~\cite{ValentinaPnch}. Events with large \ns in the \inelgz class will be characterized by a midrapidity multiplicity distribution which is biased towards higher values with respect to the unbiased one, reflected by a different \avdndeta. Conversely, events with low \ns are associated to a charged-particle-multiplicity distribution which is moved towards lower values. 
For this reason, the data points corresponding to the \inelgz class in Fig.~\ref{fig:Distributions} (a) are closer to those corresponding to the higher \VZEROM multiplicity classes X-XI in the left side of the distribution, while they lean towards the mid-multiplicity class VI in the right-hand tail.
Even when restricting to narrower \VZEROM classes, the corresponding \dndeta distribution at midrapidity is characterized by a large relative RMS, as discussed in Ref.~\cite{ANSilvia}. Therefore, this ``multiplicity bias'' effect is also present, with a different magnitude, when considering the multiplicity-differential analysis.
To quantify this effect, a MC study has been performed making use of several MC event generators. In Fig.~\ref{fig:multbias}, the results of this study are reported for the \pecrr model. 

In the left panel of Fig.~\ref{fig:multbias}, the unbiased \dndeta distributions corresponding to the \inelgz, \VZEROM-(I+II+III) and
\VZEROM-XII event classes are compared with the corresponding distributions when the requirement of having at least 5 \kzero per event is imposed. From this comparison, one can appreciate that the biased and unbiased distributions are qualitatively similar, but shifted on average (dashed lines). The magnitude of the bias depends on the number of strange particles requested and on the strange-particle species under study. This is shown on the right side of the same figure, where the relative shift in terms of \avdndeta is shown as a function of the unbiased \avdndeta in all the \VZEROM classes considered and for different biasing requirements in terms of \ns. Full markers correspond to a minimal bias ($\ns \geq 1$), while open markers are associated to a more extreme bias ($\ns \geq 5, 4, 3, 2$ for \kzero, \lmb, \X and \Om, respectively, reflecting the available statistics).
The outcome is that the bias depends on the requirement on \ns, being the highest for large \ns. Moreover, the bias depends on the strange particle considered, with a larger effect for multi-strange hadrons, as testified by the fact that requiring at least 3 \X induces a similar relative bias as requiring at least 5 \kzero. It is also worth noting that the relative shift in \avdndeta is more important in low-multiplicity classes, while it reaches values lower than 50\% for all particles and \ns at the highest multiplicity: this makes the absolute increase in average multiplicity not larger than 4 and 6 in the lowest and highest V0M classes, respectively.

Qualitatively, similar results have been obtained with \eplhc and \pem, but with a magnitude that can vary in the different models. As an example, in the lowest multiplicity class requiring $n_{\lmb}\geq4$ induces a relative \avdndeta shift of 320\%, 215\% and 500\% in the \pecrr, \pem and \eplhc cases, respectively. This difference does not come as a surprise, as it reflects the different correlations between strange and non-strange hadron formation arising from the microscopic production mechanisms in the phenomenological models. In a sense, the strangeness-induced charged-particle multiplicity bias is a complementary way of looking into the strangeness-enhancement puzzle: in a model that correctly reproduces the bias, strangeness-enhancement arises automatically correct. 

This multiplicity bias can pose a potential issue when looking into particle ratios as a function of the multiplicity. As an example, when the \OmNosign/$\pi$ ratio is computed, as in Ref.~\cite{NP}, the average charged-particle multiplicity of events containing at least one \OmNosign is 50\% higher than the unbiased average (see Fig.~\ref{fig:multbias}), so that the ratio itself is evaluated at two slightly different multiplicities. The effect is even more important if, e.g., the ratio of production probabilities for multi-strange multiplets with respect to single \kzero is calculated, as the multiplicity bias for the numerator and denominator can be very different. Despite this, when comparing results to MC generators the exact multiplicity position where the ratio is shown for the given \VZEROM class is not an essential information, as the degree of agreement/disagreement would be completely determined by the ability of the model in reproducing the relative production probabilities of particles containing different amounts of strange quarks. 
For this reason, particle ratios as a function of the multiplicity will be shown at the \avdndeta measured in the unbiased sample, named \dndetaNew. This has to be regarded as the value of the ratio that is measured when a set of collisions corresponding to a given \VZEROM class is analyzed. This is consistent with what has been previously done in all ALICE publications when yield ratios to pions have been presented.

\subsection{Multiple strange-hadron production yields}
From the measurement of P(\ns), Eq.~\ref{eq:yields} permits one to calculate the average yield \ynp{n}{S} for the production of $n$ particles of type S per event, namely the $\langle \dndy \rangle$ when $n = 1$ (mean of the strange-particle-multiplicity distribution), and all the multiple production yields (from particle pairs up to where the P(\ns) measurement extends). 

\begin{figure}[ht!]
    \begin{center}
    \includegraphics[width=0.95 \textwidth]{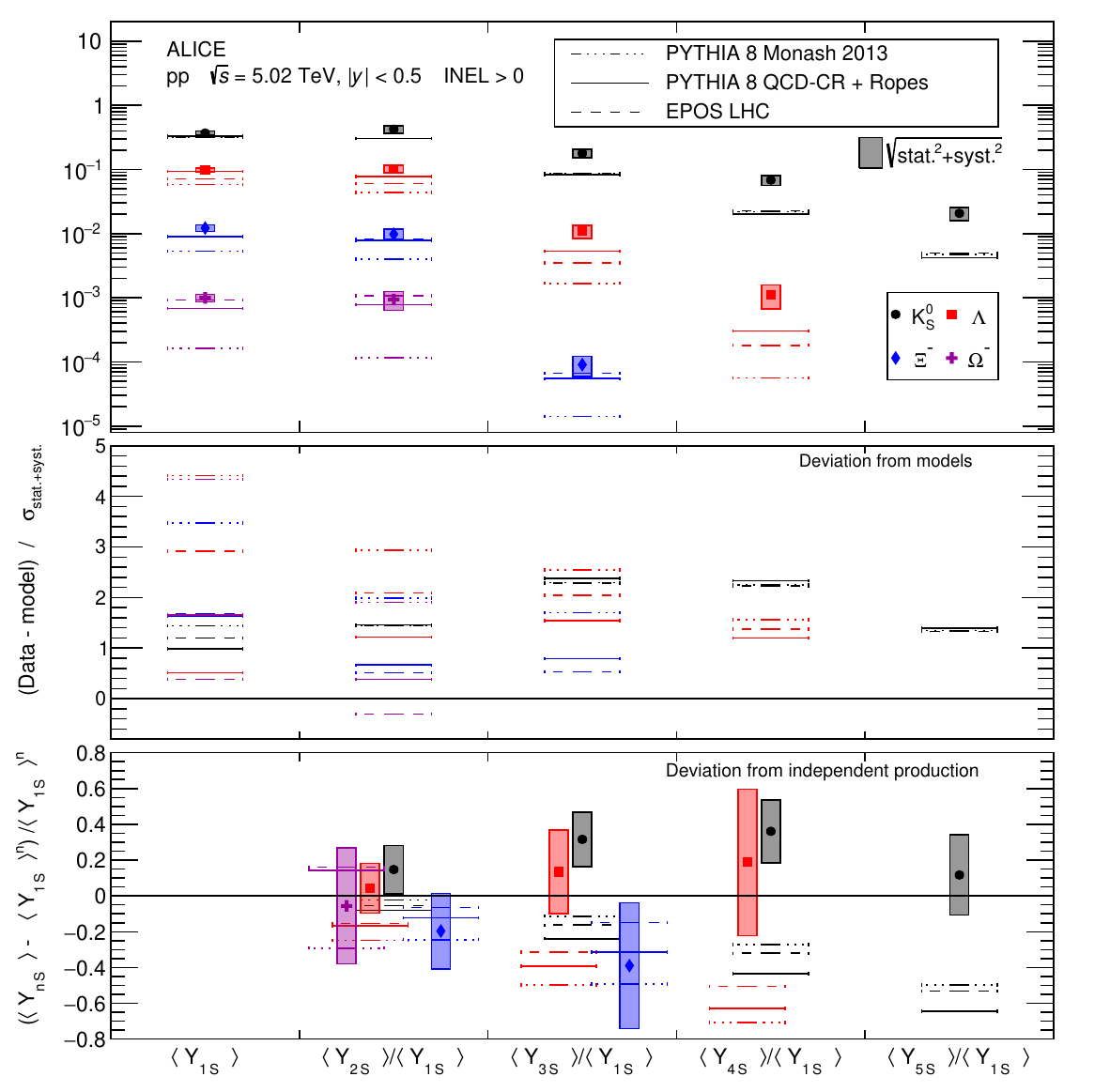}
    \end{center}
    \caption{Top panel: \ynp{1}{S} and \ynp{n>1}{S}/\ynp{1}{S} in the \inelgz event class for all particles under study. Model predictions by \pem, \pecrr and \eplhc are shown as dotted, continuous, and dashed lines, respectively. Central panel: deviation between data and models in units of the total uncertainty. Bottom panel: deviation from independent production hypothesis for both data and models. Results for different particles are shifted in $x$ to improve visibility.}
    \label{fig:RatiosMBmodel}
\end{figure}

The upper plot of Fig.~\ref{fig:RatiosMBmodel} shows \ynp{1}{S} and several average yield ratios to \ynp{1}{S} for all the hadrons under study in the \inelgz event class. The ratio makes it possible to test the ability of MC generators to reproduce the strange-particle-multiplicity distribution beyond the first moment, since the normalization to \ynp{1}{S} factors out potential incompatibilities in the average strange-particle production rate.
The central panel of the figure shows the difference between data and MC normalized to the quadratic sum of statistical and systematic uncertainty from the data.
The compatibility of \pecrr and \eplhc to the average of the distribution (first column) is within 2-3$\sigma$ for all particles, while \pem shows a larger discrepancy, testifying the importance of the improvements in the PYTHIA generator with the introduction of QCD-inspired CR and color-ropes hadronization. The impact of these improvements is evident when considering the baryons, while the \kzero yield remains practically unchanged. The comparison of the ratios \ynp{n>1}{S}/\ynp{1}{S} to model predictions shows disagreements not larger than 2.5$\sigma$ across $n$ and particle species. \newline
The bottom panel of the figure shows the deviation from the independent production hypothesis (i.e. the $n$-order multiplet yield is the $n$-th power of the single particle yield: \ynp{n}{S} = \ynp{1}{S}$^{n}$) for both data and models. Data do not exhibit a significant deviation from independent production, although the large systematic uncertainties make it difficult to draw strong conclusions. On the other hand, models tend to underestimate the production at high $n$ relative to the independent assumption. 

\begin{figure}[ht!]
    \begin{center}
    \includegraphics[width = 0.97\textwidth]{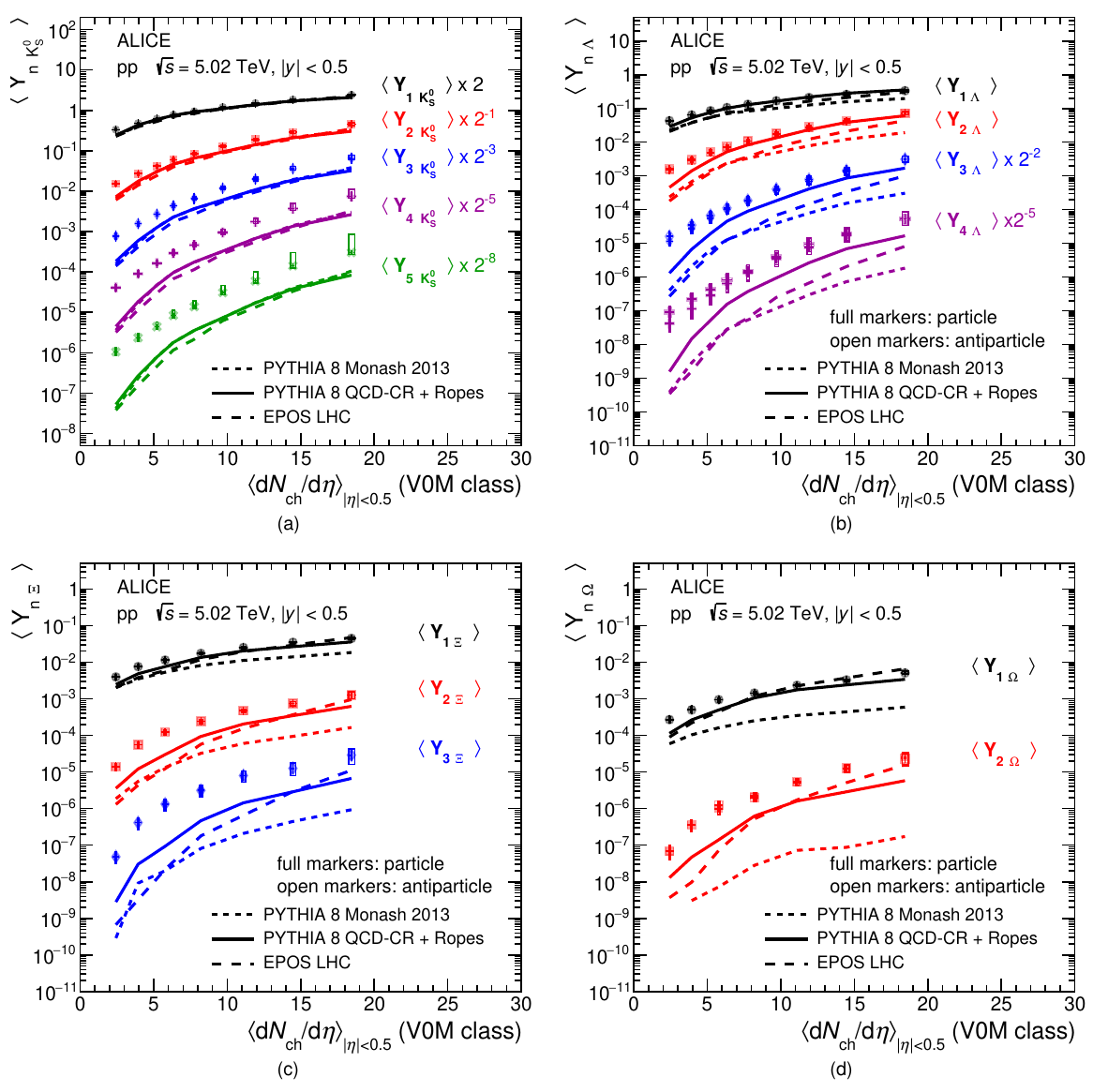}
    \end{center}
    \caption{Multiple strange-hadron production yields as a function of the unbiased average charged-particle multiplicity \dndetaNew for \kzero (a), \lmb (b), \XiNosign (c) and \OmNosign (d). Full markers correspond to particles, open markers to antiparticles. Model predictions are reported for \pem, \pecrr and \eplhc as dotted, continuous and dashed lines respectively.}
    \label{fig:yields}
\end{figure}

Multiple production yields are also calculated in different \VZEROM multiplicity classes, and the results are reported in Fig.~\ref{fig:yields} for particles (full markers) and antiparticles (open markers).
As discussed in Sec.~\ref{par:SICPMB}, the $x$-axis position of the data points is here defined as the average of the unbiased charged-particle-multiplicity distribution in the \VZEROM class, while for multiple strange-particle production, a bias can be expected that moves the averages to larger values. It is worth noting, however, that this bias would make the enhancement of the yields with multiplicity even more prominent, as the largest effect is expected at low multiplicity (see Fig.~\ref{fig:multbias}).

As already mentioned, \ynp{1}{S} corresponds to the usual $\langle \dndy \rangle$ progression with multiplicity, as previously measured. Indeed, the comparison of these results with those reported in Ref.~\cite{ANSilvia} shows perfect agreement, further demonstrating the solidity of this novel technique for strange-particle yield measurements. 
The rise with multiplicity is more than linear for the production of multiple strange particles and it becomes steeper as the number of produced hadrons increases. \ynp{5}{\kzero}, \ynp{4}{\lmb}, \ynp{3}{\XiNosign} and \ynp{2}{\OmNosign} have an increase of $\sim 2-3$ orders of magnitude from the lowest to the highest charged-particle multiplicity classes.
In Fig.~\ref{fig:yields}, the comparison with phenomenological models is also displayed. \pem has the poorest agreement with the data, with discrepancies that are significant even for \ynp{1}{S} for baryons and that get larger as the number of strange hadrons considered increases, reaching 2 orders of magnitude in the worst cases. The situation improves significantly with the \pecrr model for baryons, but it remains unchanged for \kzero, demonstrating that strange-quark production is still an open issue in PYTHIA. \eplhc has a better agreement with the data at high-multiplicity for baryons, but it undershoots the trend at low multiplicity, suggesting that the model works better in the region where the in-medium hadronization dominates. For \kzero, no clear difference among the three tested models is visible.

Strangeness-enhancement is observed when taking the ratio between production yields of particles with different strangeness content. The reference can be pions (as in Ref.~\cite{NP}) or hadrons with a single strange quark like \kzero (as in Ref.~\cite{pp13}). In previous works, the maximum unbalance of strange-quark content (\dels) between numerator and denominator was limited to three, as the ratio was performed between yields of single particles and the maximum strangeness content of a single hadron is three for the \OmNosign.

\begin{figure}[ht!]
    \begin{center}
    \includegraphics[width=0.6\textwidth]{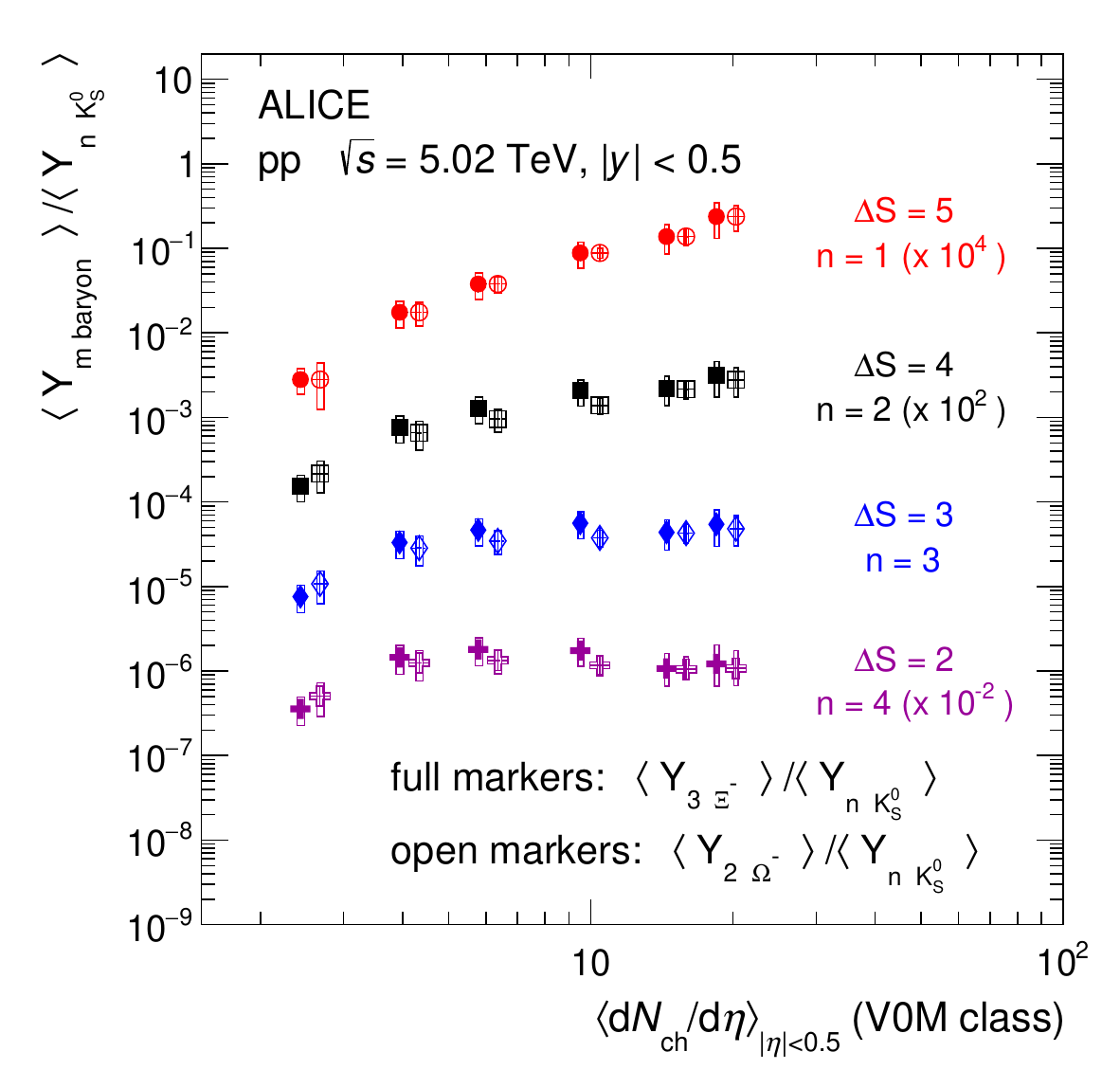}
    \end{center}
    \caption{Ratios between the average production yield of $m$ baryons, \ynp{m\,}{baryon}, and $n$ \kzero, \ynp{n\,}{\kzero}, as a function of the unbiased average charged-particle multiplicity, \dndetaNew. The case $m=3$, corresponding to \ynp{3}{\X}, is illustrated with full markers, while $m=2$, corresponding to \ynp{2}{\Om}, is shown with open markers. The value of $n$ ranges from 1 to 4, from top to bottom, respectively. Open markers are shifted to the right by 10\% to enhance visibility.}
    \label{fig:SEextremes}
\end{figure}

The measurement of multiple strange-hadron production yields opens the possibility to study strangeness-enhancement with larger \dels. This is reported in Fig.~\ref{fig:SEextremes}, where the ratios \ynp{3}{\X}/\ynp{n}{\kzero} and \ynp{2}{\Om}/\ynp{n}{\kzero} are shown as a function of the charged-particle multiplicity.
For the first time, the maximum \dels achieved is five, which corresponds to an unprecedented enhancement of two orders of magnitude from the lowest to the highest \VZEROM class. Smaller \dels are also shown, leading to progressively lower enhancement factors, as expected. The relative increases observed for \dels = 3 and 2 are compatible, within systematic uncertainties, to those previously observed in the $\XiNosign/\pi$ and $\OmNosign/\pi$ ratios ~\cite{NP}.

\begin{figure}[ht!]
    \begin{center}
    \includegraphics[width=0.99 \textwidth]{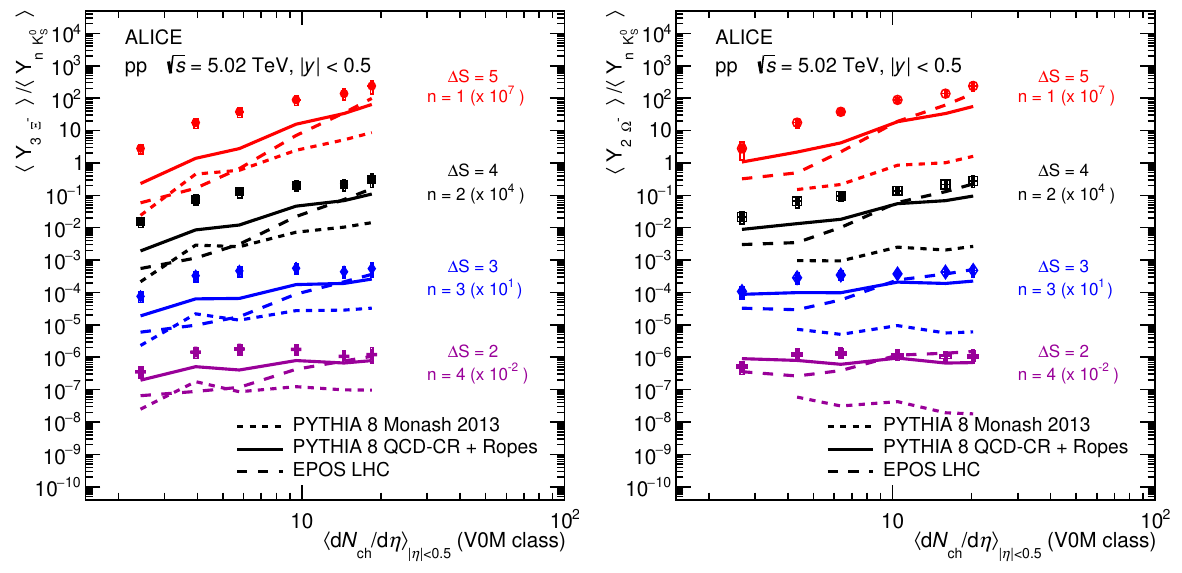}
    \end{center}
    \caption{\ynp{3}{\X}/\ynp{n}{\kzero} (\textit{left}) and \ynp{2}{\Om}/\ynp{n}{\kzero} (\textit{right}) as a function of the unbiased charged-particle multiplicity density compared with MC generators: \pem (dotted line), \pecrr (continuous line) and \eplhc (dashed line).}
    \label{fig:SEextremes_MC}
\end{figure}

It is worth noting that the trends observed with \ynp{3}{\X} and \ynp{2}{\Om} are compatible, both in terms of relative increase and in absolute value (note that the \ynp{2}{\Om} ratios are shifted to the right by 10\% to enhance visibility). This observation, which follows from the agreement between \ynp{3}{\X} and \ynp{2}{\Om} in the bottom plots of Fig.~\ref{fig:yields}, indicates that the enhancement pattern is mostly determined by the strangeness content in the hadron, rather than by other factors such as the number of baryons that have to be formed, the overall mass to be created or the number of light quarks involved. More considerations on this will be done later by comparing ratios with \dels = 0.\\ 
The comparison with MC generators is performed in Fig.~\ref{fig:SEextremes_MC}. 
\eplhc shows a very good agreement with the data at the highest \dndetaNew, but the trend with multiplicity does not match the one observed, underestimating the ratio by two orders of magnitude at very low multiplicity in the \dels = 5 case. This indicates that relative strangeness production in the core, modeled as statistical production, is well described while the transition to a corona-dominated system at low multiplicity tends to severely underestimate the production of multiple multi-strange baryons.
\pecrr shows a reasonable agreement with the data at high multiplicity, while also in this case the multiplicity progression departs from the measured trend, underestimating the data at low multiplicity, though less severely than \eplhc.
It is worth noting that, despite the observed consistency between \ynp{3}{\X}/\ynp{n}{\kzero} and \ynp{2}{\Om}/\ynp{n}{\kzero} in the data, both MC generators have different multiplicity dependencies, with a steeper increase with multiplicity when a larger number of baryons is involved (\ynp{3}{\X}). This demonstrates that the multiplicity increase in these two models is affected by the different final states in the numerator (3\X or 2\Om), a feature that is not observed in the data.

\begin{figure}[ht!]
    \begin{center}
    \includegraphics[width=0.6 \textwidth]{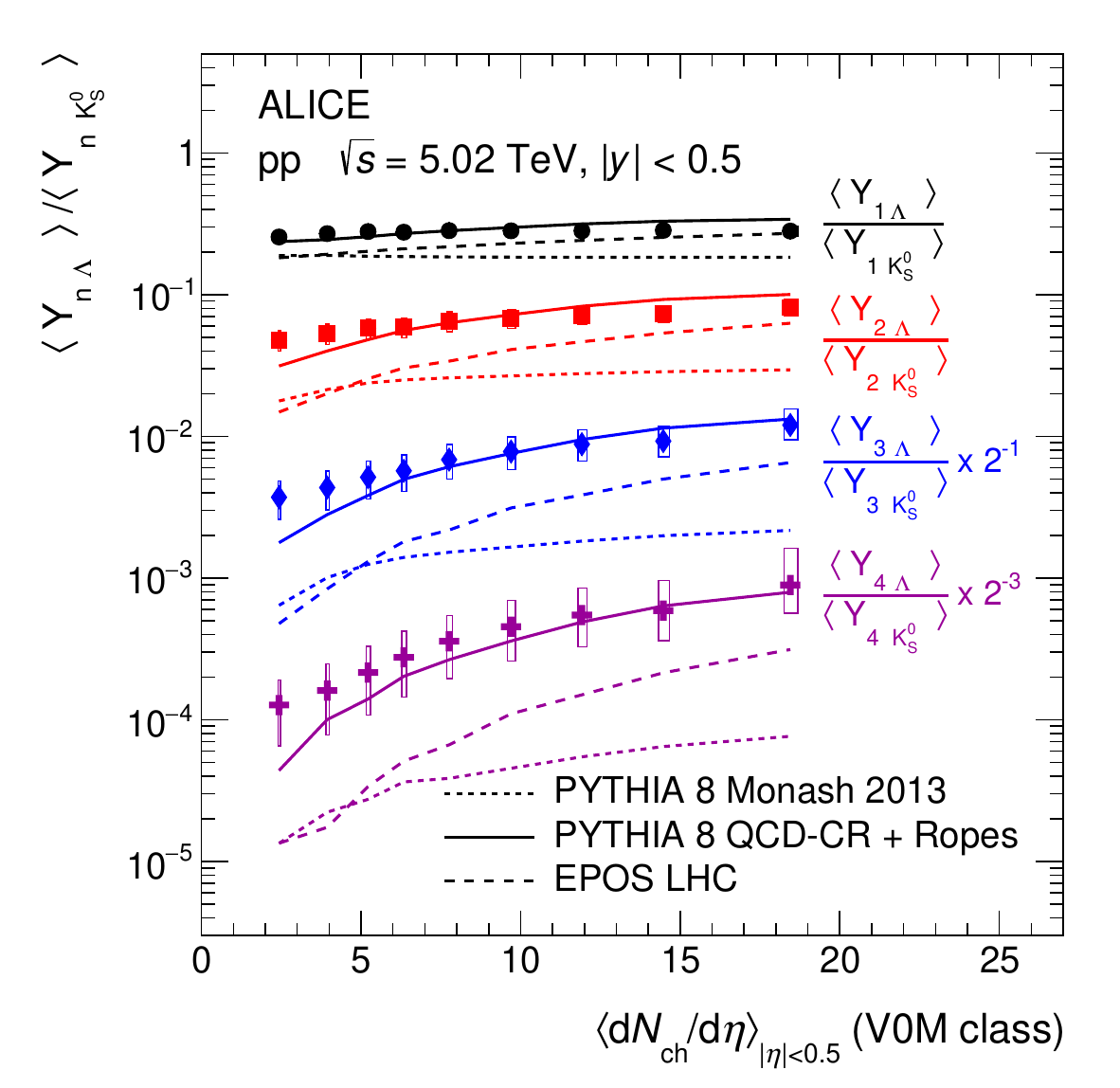}
    \end{center}
    \caption{\ynp{n}{\lmb}/\ynp{n}{\kzero} as a function of the unbiased average charged-particle multiplicity at midrapidity. $n$ runs from 1 to 4, from top to bottom, respectively. MC predictions from \pem (dotted lines), \pecrr (continuous lines) and \eplhc (dashed lines) are also shown.}
    \label{fig:lamovk0}
\end{figure}

\begin{figure}[ht!]
    \begin{center}
    \includegraphics[width=0.98 \textwidth]{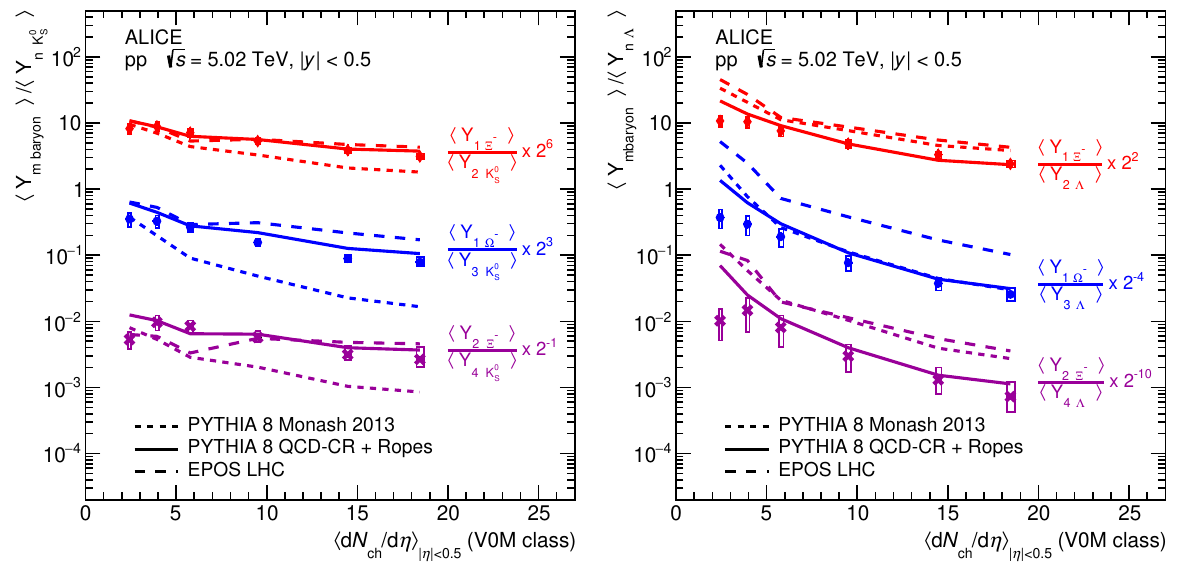}
    \end{center}
    \caption{Ratios of the average production yield of $m$ multi-strange baryons \ynp{m\,}{baryon}, to those of $n$ single-strange particles, \ynp{n\,}{\kzero} (\textit{left}) and \ynp{n\,}{\lmb} (\textit{right}), as a function of the unbiased average charged-particle multiplicity, \dndetaNew. The measurements are compared to MC predictions from \pem (dotted lines), \pecrr (continuous lines) and \eplhc (dashed lines). (\textit{left}) \ynp{1}{\X}/\ynp{2}{\kzero}, \ynp{1}{\Om}/\ynp{3}{\kzero} and \ynp{2}{\X}/\ynp{4}{\kzero} from top to bottom respectively. (\textit{right}) \ynp{1}{\X}/\ynp{2}{\lmb}, \ynp{1}{\Om}/\ynp{3}{\lmb} and \ynp{2}{\X}/\ynp{4}{\lmb} from top to bottom respectively.}
    \label{fig:SBeffect}
\end{figure}

The enhancement pattern in the strange-to-pion ratio as a function of the multiplicity is generally attributed to the strangeness content in the hadron. The argument is that such an effect was not observed in ratios with equal amount of strange quarks in the numerator and in the denominator. It is the case for \lmb/\kzero and p/$\pi$ that there is no clear enhancement pattern in the midrapidity multiplicity range between $\sim5$ and $\sim50$~\cite{NP}. For lower multiplicity, a significant depletion of these ratios was observed, but limited to a single data point in each case. 
To further test the potential presence of a non strangeness-related effect in the evolution of particle ratios with multiplicity, the relative production of different multiplets with \dels = 0 are computed and reported in Figs.~\ref{fig:lamovk0} and ~\ref{fig:SBeffect}.

In Fig.~\ref{fig:lamovk0}, for the first time the \ynp{n}{\lmb}/\ynp{n}{\kzero} ratio shows an increase with \dndetaNew for $n>1$, demonstrating that the evolution with multiplicity of multi-particle yield ratios is not solely connected to the strangeness imbalance between numerator and denominator. For $n=1$ the ratio is consistent with what was reported in Refs.~\cite{NP,pp13}, confirming that the observed enhancement at \dels = 0  is sub-leading with respect to the strangeness-related effect. Whether the \ynp{n}{\lmb}/\ynp{n}{\kzero} enhancement with multiplicity is connected to a baryon effect, to the larger total mass in the numerator, to the presence of \kzero pairs with zero net strangeness, most prominent at low multiplicity, or to the amount of $u$ and $d$ quarks required to form a \lmb or a \kzero cannot be concluded from Fig.~\ref{fig:lamovk0} alone. 

Figure~\ref{fig:SBeffect} presents additional examples of \dels = 0 multiplet ratios as a function of the charged particle multiplicity. In the left panel, \ynp{1}{\X}/\ynp{2}{\kzero}, \ynp{1}{\Om}/\ynp{3}{\kzero}, and \ynp{2}{\X}/\ynp{4}{\kzero} are reported. All these combinations feature a higher overall mass and a higher baryon content in the numerator, but the ratios decrease with multiplicity.
This suggests that mass and baryon number are not directly correlated with an enhancement as multiplicity increases. On the other hand, within a na\"ive quark coalescence scenario, the trends reported in Fig.~\ref{fig:lamovk0} and the left panel of Fig.~\ref{fig:SBeffect} can be explained. Taking the \ynp{1}{\Om}/\ynp{3}{\kzero} ratio as an example, the requirement of an \Om or three \kzero fixes the minimum number of strange quarks to 3. However, forming an \Om requires no additional $u$ or $d$, while forming three \kzero requires three light quarks. In a low-multiplicity environment, the scarcity of light quarks increases the likelihood that the three strange quarks will form an \Om rather than three \kzero. Conversely, in a high-multiplicity environment, light quarks are abundant, making it easier to pair them with a strange one. Note that quark coalescence in small collision systems has been recently proposed to describe the distinctive grouping of identified hadron $v_2$ based on their valence quark number~\cite{alicecollaboration2024observationpartonicflowprotonproton}.  
A qualitatively similar pattern is observed in the right panel of Fig.~\ref{fig:SBeffect}, which shows the ratios \ynp{1}{\X}/\ynp{2}{\lmb}, \ynp{1}{\Om}/\ynp{3}{\lmb}  and \ynp{2}{\X}/\ynp{4}{\lmb}. Quantitatively, when ratios to \kzero or \lmb are compared fixing the number of \XiNosign or \OmNosign in the numerator, the decreasing trend with multiplicity is more pronounced when a baryon is present in the denominator. This can again be interpreted considering that a \lmb baryon contains two light quarks, while a \kzero meson only one. Despite the simplicity of this explanation, the qualitative idea provides a consistent description for all the ratios shown in Fig.~\ref{fig:lamovk0} and Fig.~\ref{fig:SBeffect}.

Model comparisons in this context are very enlightening. \pem underestimates all particle ratios to \kzero (Fig.~\ref{fig:lamovk0} and the left panel of Fig.~\ref{fig:SBeffect}), while it overestimates ratios to \lmb (right panel of Fig.~\ref{fig:SBeffect}), with the only exception of \ynp{1}{\Om}/\ynp{3}{\lmb}. Additionally, it fails to reproduce the magnitude of the observed increasing or decreasing trends. In contrast, \pecrr provides an excellent description of all trends, both qualitatively and quantitatively, demonstrating that the QCD-CR mechanism implemented in the model effectively captures the reconnection of strange and light quarks as a function of multiplicity. \eplhc reasonably explains the multiplicity trend for all ratios to \kzero, while when the ratio is performed with respect to \lmb production, it consistently overestimates the observed values.


\section{Summary} \label{conclusion}
The production probability of multiple (multi-)strange hadrons, \pns, has been studied at midrapidity analyzing \five proton-proton collisions, dividing events into several charged-particle multiplicity-density classes. 
The analysis technique, used for the first time for strange particle yields extraction, is based on counting the number of strange particles event-by-event and then correcting the raw multiplicity distribution with a one-dimensional Bayesian unfolding procedure. This technique allows one to extend the study of strangeness production beyond the average of the multiplicity distribution and constitutes a new test bench for MC generators, probing events with a large imbalance between the number of strange and non-strange hadrons produced.\\
The probability of producing more than one strange-particle of a given species per event increases more than linearly with the event charged-particle multiplicity. When high \ns are accessed, \pns in the \inelgz class shows a tail which approaches the probabilities measured in high-multiplicity \VZEROM classes. This feature originates from the bias in the charged-particle-multiplicity distribution induced by the request of having a large number of strange particles in the event. The study of this bias in MC generators is an important tool to explore the connection between strange and non-strange-particle production.\\
From the measurement of \pns, the average multiplet production yields \ynp{n}{S} have been calculated in every \VZEROM multiplicity class and for all particles under study. 
An enhancement with multiplicity is observed, with a steeper increase for higher $n$. The comparison with the PYTHIA MC generator shows improvements achieved in the \pecrr implementation of the model for baryons, while no specific improvement is observed for \kzero. The \eplhc model has a better agreement with the data at high multiplicity, while it severely underestimates \ynp{n}{S} at low multiplicity.\\
Strangeness-enhancement has been shown with unprecedented magnitude by evaluating the ratio of cascade multiplets to single \kzero production, thus reaching strange content imbalances between numerator and denominator up to \dels = 5. Comparing \ynp{3}{\Xi}/\ynp{n}{\kzero} and \ynp{3}{\Om}/\ynp{n}{\kzero} a perfect agreement is observed, demonstrating that the enhancement is dominated by the strangeness-content imbalance. Nonetheless, multiplet ratios with \dels = 0 showcase an evolution with multiplicity, which demonstrates that a sub-leading role is played by light $u$ and $d$ quarks as well. In particular, \ynp{n}{\lmb}/\ynp{n}{\kzero} shows for the first time an enhancement pattern with multiplicity for large $n$, while in the \ynp{m\,}{baryon}/\ynp{n}{\kzero} and \ynp{m\,}{baryon}/\ynp{n}{\lmb} ratios a decreasing trend is observed. These features can be reconciled in a basic coalescence scenario, where the probability to associate a light $u$ or $d$ quark to a $s$ quark evolves with the event multiplicity. All the \dels = 0 ratios are well reproduced by the \pecrr model, further supporting the specific QCD-CR mechanism there implemented for strange baryons.\\
The results reported in this paper pave the way to a new generation of strange-particle yield measurements, allowing studies of the relative probability, for a given number of strange quarks produced in a \pp event, to hadronize in different combinations of strange hadrons in the final state. 
Larger statistics acquired during the LHC Run~3 campaign will allow a significant extension of the measurements to higher strange-particle multiplicities and, potentially, to study the simultaneous production of different strange-hadron species.


\newenvironment{acknowledgement}{\relax}{\relax}
\begin{acknowledgement}
\section*{Acknowledgements}

The ALICE Collaboration would like to thank all its engineers and technicians for their invaluable contributions to the construction of the experiment and the CERN accelerator teams for the outstanding performance of the LHC complex.
The ALICE Collaboration gratefully acknowledges the resources and support provided by all Grid centres and the Worldwide LHC Computing Grid (WLCG) collaboration.
The ALICE Collaboration acknowledges the following funding agencies for their support in building and running the ALICE detector:
A. I. Alikhanyan National Science Laboratory (Yerevan Physics Institute) Foundation (ANSL), State Committee of Science and World Federation of Scientists (WFS), Armenia;
Austrian Academy of Sciences, Austrian Science Fund (FWF): [M 2467-N36] and Nationalstiftung f\"{u}r Forschung, Technologie und Entwicklung, Austria;
Ministry of Communications and High Technologies, National Nuclear Research Center, Azerbaijan;
Rede Nacional de Física de Altas Energias (Renafae), Financiadora de Estudos e Projetos (Finep), Funda\c{c}\~{a}o de Amparo \`{a} Pesquisa do Estado de S\~{a}o Paulo (FAPESP) and The Sao Paulo Research Foundation  (FAPESP), Brazil;
Bulgarian Ministry of Education and Science, within the National Roadmap for Research Infrastructures 2020-2027 (object CERN), Bulgaria;
Ministry of Education of China (MOEC) , Ministry of Science \& Technology of China (MSTC) and National Natural Science Foundation of China (NSFC), China;
Ministry of Science and Education and Croatian Science Foundation, Croatia;
Centro de Aplicaciones Tecnol\'{o}gicas y Desarrollo Nuclear (CEADEN), Cubaenerg\'{\i}a, Cuba;
Ministry of Education, Youth and Sports of the Czech Republic, Czech Republic;
The Danish Council for Independent Research | Natural Sciences, the VILLUM FONDEN and Danish National Research Foundation (DNRF), Denmark;
Helsinki Institute of Physics (HIP), Finland;
Commissariat \`{a} l'Energie Atomique (CEA) and Institut National de Physique Nucl\'{e}aire et de Physique des Particules (IN2P3) and Centre National de la Recherche Scientifique (CNRS), France;
Bundesministerium f\"{u}r Forschung, Technologie und Raumfahrt (BMFTR) and GSI Helmholtzzentrum f\"{u}r Schwerionenforschung GmbH, Germany;
General Secretariat for Research and Technology, Ministry of Education, Research and Religions, Greece;
National Research, Development and Innovation Office, Hungary;
Department of Atomic Energy Government of India (DAE), Department of Science and Technology, Government of India (DST), University Grants Commission, Government of India (UGC) and Council of Scientific and Industrial Research (CSIR), India;
National Research and Innovation Agency - BRIN, Indonesia;
Istituto Nazionale di Fisica Nucleare (INFN), Italy;
Japanese Ministry of Education, Culture, Sports, Science and Technology (MEXT) and Japan Society for the Promotion of Science (JSPS) KAKENHI, Japan;
Consejo Nacional de Ciencia (CONACYT) y Tecnolog\'{i}a, through Fondo de Cooperaci\'{o}n Internacional en Ciencia y Tecnolog\'{i}a (FONCICYT) and Direcci\'{o}n General de Asuntos del Personal Academico (DGAPA), Mexico;
Nederlandse Organisatie voor Wetenschappelijk Onderzoek (NWO), Netherlands;
The Research Council of Norway, Norway;
Pontificia Universidad Cat\'{o}lica del Per\'{u}, Peru;
Ministry of Science and Higher Education, National Science Centre and WUT ID-UB, Poland;
Korea Institute of Science and Technology Information and National Research Foundation of Korea (NRF), Republic of Korea;
Ministry of Education and Scientific Research, Institute of Atomic Physics, Ministry of Research and Innovation and Institute of Atomic Physics and Universitatea Nationala de Stiinta si Tehnologie Politehnica Bucuresti, Romania;
Ministerstvo skolstva, vyskumu, vyvoja a mladeze SR, Slovakia;
National Research Foundation of South Africa, South Africa;
Swedish Research Council (VR) and Knut \& Alice Wallenberg Foundation (KAW), Sweden;
European Organization for Nuclear Research, Switzerland;
Suranaree University of Technology (SUT), National Science and Technology Development Agency (NSTDA) and National Science, Research and Innovation Fund (NSRF via PMU-B B05F650021), Thailand;
Turkish Energy, Nuclear and Mineral Research Agency (TENMAK), Turkey;
National Academy of  Sciences of Ukraine, Ukraine;
Science and Technology Facilities Council (STFC), United Kingdom;
National Science Foundation of the United States of America (NSF) and United States Department of Energy, Office of Nuclear Physics (DOE NP), United States of America.
In addition, individual groups or members have received support from:
Czech Science Foundation (grant no. 23-07499S), Czech Republic;
FORTE project, reg.\ no.\ CZ.02.01.01/00/22\_008/0004632, Czech Republic, co-funded by the European Union, Czech Republic;
European Research Council (grant no. 950692), European Union;
Deutsche Forschungs Gemeinschaft (DFG, German Research Foundation) ``Neutrinos and Dark Matter in Astro- and Particle Physics'' (grant no. SFB 1258), Germany;
CONVECS project, CUP C97H23001700002 FESR 2021-2027 program, Italy;
FAIR - Future Artificial Intelligence Research, funded by the NextGenerationEU program (Italy).

\end{acknowledgement}

\bibliographystyle{utphys}   
\bibliography{bibliography}

\newpage
\appendix

%
%

\section{The ALICE Collaboration}
\label{app:collab}
\begin{flushleft} 
\small

I.J.~Abualrob\,\orcidlink{0009-0005-3519-5631}\,$^{\rm 114}$, 
S.~Acharya\,\orcidlink{0000-0002-9213-5329}\,$^{\rm 50}$, 
G.~Aglieri Rinella\,\orcidlink{0000-0002-9611-3696}\,$^{\rm 32}$, 
L.~Aglietta\,\orcidlink{0009-0003-0763-6802}\,$^{\rm 24}$, 
N.~Agrawal\,\orcidlink{0000-0003-0348-9836}\,$^{\rm 25}$, 
Z.~Ahammed\,\orcidlink{0000-0001-5241-7412}\,$^{\rm 134}$, 
S.~Ahmad\,\orcidlink{0000-0003-0497-5705}\,$^{\rm 15}$, 
I.~Ahuja\,\orcidlink{0000-0002-4417-1392}\,$^{\rm 36}$, 
Z.~Akbar$^{\rm 81}$, 
A.~Akindinov\,\orcidlink{0000-0002-7388-3022}\,$^{\rm 140}$
% ...

\section*{Affiliation Notes}

$^{\rm I}$ Deceased\\
$^{\rm II}$ Also at: Max-Planck-Institut fur Physik, Munich, Germany\\
$^{\rm III}$ Also at: Czech Technical University in Prague, Prague, Czech Republic\\
$^{\rm IV}$ Also at: Instituto de Fisica da Universidade de Sao Paulo\\
$^{\rm V}$ Also at: Dipartimento DET del Politecnico di Torino, Turin, Italy\\
$^{\rm VI}$ Also at: University College of Dublin, Dublin, Ireland\\
$^{\rm VII}$ Also at: Department of Applied Physics, Aligarh Muslim University, Aligarh, India\\
$^{\rm VIII}$ Also at: Institute of Theoretical Physics, University of Wroclaw, Wroclaw, Poland\\
$^{\rm IX}$ Also at: Facultad de Ciencias, Universidad Nacional Aut\'{o}noma de M\'{e}xico, Mexico City, Mexico\\

\section*{Collaboration Institutes}

$^{1}$ A.I. Alikhanyan National Science Laboratory (Yerevan Physics Institute) Foundation, Yerevan, Armenia\\
$^{2}$ AGH University of Krakow, Cracow, Poland\\
$^{3}$ Bogolyubov Institute for Theoretical Physics, National Academy of Sciences of Ukraine, Kyiv, Ukraine\\
$^{4}$ Bose Institute, Department of Physics  and Centre for Astroparticle Physics and Space Science (CAPSS), Kolkata, India\\
$^{5}$ California Polytechnic State University, San Luis Obispo, California, United States\\
$^{6}$ Central China Normal University, Wuhan, China\\
$^{7}$ Centro de Aplicaciones Tecnol\'{o}gicas y Desarrollo Nuclear (CEADEN), Havana, Cuba\\
$^{8}$ Centro de Investigaci\'{o}n y de Estudios Avanzados (CINVESTAV), Mexico City and M\'{e}rida, Mexico\\
$^{9}$ Chicago State University, Chicago, Illinois, United States\\
$^{10}$ China Nuclear Data Center, China Institute of Atomic Energy, Beijing, China\\
$^{11}$ China University of Geosciences, Wuhan, China\\
$^{12}$ Chungbuk National University, Cheongju, Republic of Korea\\
$^{13}$ Comenius University Bratislava, Faculty of Mathematics, Physics and Informatics, Bratislava, Slovak Republic\\
$^{14}$ Creighton University, Omaha, Nebraska, United States\\
$^{15}$ Department of Physics, Aligarh Muslim University, Aligarh, India\\
$^{16}$ Department of Physics, Pusan National University, Pusan, Republic of Korea\\
$^{17}$ Department of Physics, Sejong University, Seoul, Republic of Korea\\
$^{18}$ Department of Physics, University of California, Berkeley, California, United States\\
$^{19}$ Department of Physics, University of Oslo, Oslo, Norway\\
$^{20}$ Department of Physics and Technology, University of Bergen, Bergen, Norway\\
$^{21}$ Dipartimento di Fisica, Universit\`{a} di Pavia, Pavia, Italy\\
$^{22}$ Dipartimento di Fisica dell'Universit\`{a} and Sezione INFN, Cagliari, Italy\\
$^{23}$ Dipartimento di Fisica dell'Universit\`{a} and Sezione INFN, Trieste, Italy\\
$^{24}$ Dipartimento di Fisica dell'Universit\`{a} and Sezione INFN, Turin, Italy\\
$^{25}$ Dipartimento di Fisica e Astronomia dell'Universit\`{a} and Sezione INFN, Bologna, Italy\\
$^{26}$ Dipartimento di Fisica e Astronomia dell'Universit\`{a} and Sezione INFN, Catania, Italy\\
$^{27}$ Dipartimento di Fisica e Astronomia dell'Universit\`{a} and Sezione INFN, Padova, Italy\\
$^{28}$ Dipartimento di Fisica `E.R.~Caianiello' dell'Universit\`{a} and Gruppo Collegato INFN, Salerno, Italy\\
$^{29}$ Dipartimento DISAT del Politecnico and Sezione INFN, Turin, Italy\\
$^{30}$ Dipartimento di Scienze MIFT, Universit\`{a} di Messina, Messina, Italy\\
$^{31}$ Dipartimento Interateneo di Fisica `M.~Merlin' and Sezione INFN, Bari, Italy\\
$^{32}$ European Organization for Nuclear Research (CERN), Geneva, Switzerland\\
$^{33}$ Faculty of Electrical Engineering, Mechanical Engineering and Naval Architecture, University of Split, Split, Croatia\\
$^{34}$ Faculty of Nuclear Sciences and Physical Engineering, Czech Technical University in Prague, Prague, Czech Republic\\
$^{35}$ Faculty of Physics, Sofia University, Sofia, Bulgaria\\
$^{36}$ Faculty of Science, P.J.~\v{S}af\'{a}rik University, Ko\v{s}ice, Slovak Republic\\
$^{37}$ Faculty of Technology, Environmental and Social Sciences, Bergen, Norway\\
$^{38}$ Frankfurt Institute for Advanced Studies, Johann Wolfgang Goethe-Universit\"{a}t Frankfurt, Frankfurt, Germany\\
$^{39}$ Fudan University, Shanghai, China\\
$^{40}$ Gangneung-Wonju National University, Gangneung, Republic of Korea\\
$^{41}$ Gauhati University, Department of Physics, Guwahati, India\\
$^{42}$ Helmholtz-Institut f\"{u}r Strahlen- und Kernphysik, Rheinische Friedrich-Wilhelms-Universit\"{a}t Bonn, Bonn, Germany\\
$^{43}$ Helsinki Institute of Physics (HIP), Helsinki, Finland\\
$^{44}$ High Energy Physics Group,  Universidad Aut\'{o}noma de Puebla, Puebla, Mexico\\
$^{45}$ Horia Hulubei National Institute of Physics and Nuclear Engineering, Bucharest, Romania\\
$^{46}$ HUN-REN Wigner Research Centre for Physics, Budapest, Hungary\\
$^{47}$ Indian Institute of Technology Bombay (IIT), Mumbai, India\\
$^{48}$ Indian Institute of Technology Indore, Indore, India\\
$^{49}$ INFN, Laboratori Nazionali di Frascati, Frascati, Italy\\
$^{50}$ INFN, Sezione di Bari, Bari, Italy\\
$^{51}$ INFN, Sezione di Bologna, Bologna, Italy\\
$^{52}$ INFN, Sezione di Cagliari, Cagliari, Italy\\
$^{53}$ INFN, Sezione di Catania, Catania, Italy\\
$^{54}$ INFN, Sezione di Padova, Padova, Italy\\
$^{55}$ INFN, Sezione di Pavia, Pavia, Italy\\
$^{56}$ INFN, Sezione di Torino, Turin, Italy\\
$^{57}$ INFN, Sezione di Trieste, Trieste, Italy\\
$^{58}$ Inha University, Incheon, Republic of Korea\\
$^{59}$ Institute for Gravitational and Subatomic Physics (GRASP), Utrecht University/Nikhef, Utrecht, Netherlands\\
$^{60}$ Institute of Experimental Physics, Slovak Academy of Sciences, Ko\v{s}ice, Slovak Republic\\
$^{61}$ Institute of Physics, Homi Bhabha National Institute, Bhubaneswar, India\\
$^{62}$ Institute of Physics of the Czech Academy of Sciences, Prague, Czech Republic\\
$^{63}$ Institute of Space Science (ISS), Bucharest, Romania\\
$^{64}$ Institut f\"{u}r Kernphysik, Johann Wolfgang Goethe-Universit\"{a}t Frankfurt, Frankfurt, Germany\\
$^{65}$ Instituto de Ciencias Nucleares, Universidad Nacional Aut\'{o}noma de M\'{e}xico, Mexico City, Mexico\\
$^{66}$ Instituto de F\'{i}sica, Universidade Federal do Rio Grande do Sul (UFRGS), Porto Alegre, Brazil\\
$^{67}$ Instituto de F\'{\i}sica, Universidad Nacional Aut\'{o}noma de M\'{e}xico, Mexico City, Mexico\\
$^{68}$ iThemba LABS, National Research Foundation, Somerset West, South Africa\\
$^{69}$ Jeonbuk National University, Jeonju, Republic of Korea\\
$^{70}$ Korea Institute of Science and Technology Information, Daejeon, Republic of Korea\\
$^{71}$ Laboratoire de Physique Subatomique et de Cosmologie, Universit\'{e} Grenoble-Alpes, CNRS-IN2P3, Grenoble, France\\
$^{72}$ Lawrence Berkeley National Laboratory, Berkeley, California, United States\\
$^{73}$ Lund University Department of Physics, Division of Particle Physics, Lund, Sweden\\
$^{74}$ Marietta Blau Institute, Vienna, Austria\\
$^{75}$ Nagasaki Institute of Applied Science, Nagasaki, Japan\\
$^{76}$ Nara Women{'}s University (NWU), Nara, Japan\\
$^{77}$ National and Kapodistrian University of Athens, School of Science, Department of Physics , Athens, Greece\\
$^{78}$ National Centre for Nuclear Research, Warsaw, Poland\\
$^{79}$ National Institute of Science Education and Research, Homi Bhabha National Institute, Jatni, India\\
$^{80}$ National Nuclear Research Center, Baku, Azerbaijan\\
$^{81}$ National Research and Innovation Agency - BRIN, Jakarta, Indonesia\\
$^{82}$ Niels Bohr Institute, University of Copenhagen, Copenhagen, Denmark\\
$^{83}$ Nikhef, National institute for subatomic physics, Amsterdam, Netherlands\\
$^{84}$ Nuclear Physics Group, STFC Daresbury Laboratory, Daresbury, United Kingdom\\
$^{85}$ Nuclear Physics Institute of the Czech Academy of Sciences, Husinec-\v{R}e\v{z}, Czech Republic\\
$^{86}$ Oak Ridge National Laboratory, Oak Ridge, Tennessee, United States\\
$^{87}$ Ohio State University, Columbus, Ohio, United States\\
$^{88}$ Physics department, Faculty of science, University of Zagreb, Zagreb, Croatia\\
$^{89}$ Physics Department, Panjab University, Chandigarh, India\\
$^{90}$ Physics Department, University of Jammu, Jammu, India\\
$^{91}$ Physics Program and International Institute for Sustainability with Knotted Chiral Meta Matter (WPI-SKCM$^{2}$), Hiroshima University, Hiroshima, Japan\\
$^{92}$ Physikalisches Institut, Eberhard-Karls-Universit\"{a}t T\"{u}bingen, T\"{u}bingen, Germany\\
$^{93}$ Physikalisches Institut, Ruprecht-Karls-Universit\"{a}t Heidelberg, Heidelberg, Germany\\
$^{94}$ Physik Department, Technische Universit\"{a}t M\"{u}nchen, Munich, Germany\\
$^{95}$ Politecnico di Bari and Sezione INFN, Bari, Italy\\
$^{96}$ Research Division and ExtreMe Matter Institute EMMI, GSI Helmholtzzentrum f\"ur Schwerionenforschung GmbH, Darmstadt, Germany\\
$^{97}$ Saga University, Saga, Japan\\
$^{98}$ Saha Institute of Nuclear Physics, Homi Bhabha National Institute, Kolkata, India\\
$^{99}$ School of Physics and Astronomy, University of Birmingham, Birmingham, United Kingdom\\
$^{100}$ Secci\'{o}n F\'{\i}sica, Departamento de Ciencias, Pontificia Universidad Cat\'{o}lica del Per\'{u}, Lima, Peru\\
$^{101}$ SUBATECH, IMT Atlantique, Nantes Universit\'{e}, CNRS-IN2P3, Nantes, France\\
$^{102}$ Sungkyunkwan University, Suwon City, Republic of Korea\\
$^{103}$ Suranaree University of Technology, Nakhon Ratchasima, Thailand\\
$^{104}$ Technical University of Ko\v{s}ice, Ko\v{s}ice, Slovak Republic\\
$^{105}$ The Henryk Niewodniczanski Institute of Nuclear Physics, Polish Academy of Sciences, Cracow, Poland\\
$^{106}$ The University of Texas at Austin, Austin, Texas, United States\\
$^{107}$ Universidad Aut\'{o}noma de Sinaloa, Culiac\'{a}n, Mexico\\
$^{108}$ Universidade de S\~{a}o Paulo (USP), S\~{a}o Paulo, Brazil\\
$^{109}$ Universidade Estadual de Campinas (UNICAMP), Campinas, Brazil\\
$^{110}$ Universidade Federal do ABC, Santo Andre, Brazil\\
$^{111}$ Universitatea Nationala de Stiinta si Tehnologie Politehnica Bucuresti, Bucharest, Romania\\
$^{112}$ University of Cape Town, Cape Town, South Africa\\
$^{113}$ University of Derby, Derby, United Kingdom\\
$^{114}$ University of Houston, Houston, Texas, United States\\
$^{115}$ University of Jyv\"{a}skyl\"{a}, Jyv\"{a}skyl\"{a}, Finland\\
$^{116}$ University of Kansas, Lawrence, Kansas, United States\\
$^{117}$ University of Liverpool, Liverpool, United Kingdom\\
$^{118}$ University of Science and Technology of China, Hefei, China\\
$^{119}$ University of Silesia in Katowice, Katowice, Poland\\
$^{120}$ University of South-Eastern Norway, Kongsberg, Norway\\
$^{121}$ University of Tennessee, Knoxville, Tennessee, United States\\
$^{122}$ University of the Witwatersrand, Johannesburg, South Africa\\
$^{123}$ University of Tokyo, Tokyo, Japan\\
$^{124}$ University of Tsukuba, Tsukuba, Japan\\
$^{125}$ Universit\"{a}t M\"{u}nster, Institut f\"{u}r Kernphysik, M\"{u}nster, Germany\\
$^{126}$ Universit\'{e} Clermont Auvergne, CNRS/IN2P3, LPC, Clermont-Ferrand, France\\
$^{127}$ Universit\'{e} de Lyon, CNRS/IN2P3, Institut de Physique des 2 Infinis de Lyon, Lyon, France\\
$^{128}$ Universit\'{e} de Strasbourg, CNRS, IPHC UMR 7178, F-67000 Strasbourg, France, Strasbourg, France\\
$^{129}$ Universit\'{e} Paris-Saclay, Centre d'Etudes de Saclay (CEA), IRFU, D\'{e}partment de Physique Nucl\'{e}aire (DPhN), Saclay, France\\
$^{130}$ Universit\'{e}  Paris-Saclay, CNRS/IN2P3, IJCLab, Orsay, France\\
$^{131}$ Universit\`{a} degli Studi di Foggia, Foggia, Italy\\
$^{132}$ Universit\`{a} del Piemonte Orientale, Vercelli, Italy\\
$^{133}$ Universit\`{a} di Brescia, Brescia, Italy\\
$^{134}$ Variable Energy Cyclotron Centre, Homi Bhabha National Institute, Kolkata, India\\
$^{135}$ Warsaw University of Technology, Warsaw, Poland\\
$^{136}$ Wayne State University, Detroit, Michigan, United States\\
$^{137}$ Yale University, New Haven, Connecticut, United States\\
$^{138}$ Yildiz Technical University, Istanbul, Turkey\\
$^{139}$ Yonsei University, Seoul, Republic of Korea\\
$^{140}$ Affiliated with an institute formerly covered by a cooperation agreement with CERN\\
$^{141}$ Affiliated with an international laboratory covered by a cooperation agreement with CERN.\\

\end{flushleft} 
  
\end{document}